\documentclass[12pt]{article}
\usepackage[a4paper, total={7in, 10in}]{geometry}
\usepackage[parfill]{parskip}
\usepackage{physics, tensor, float, subcaption}
\usepackage{graphicx}
\graphicspath{ {Plots/} }
\usepackage{jhep-mod}
\usepackage{bm}
\usepackage{soul}
\usepackage{amssymb,amsmath,amsthm}
\usepackage{mathrsfs}
\usepackage[utf8]{inputenc}
\usepackage{enumerate}
\usepackage{bigints}
\usepackage{xcolor}
\usepackage{appendix}
\usepackage{graphicx}
\usepackage{float}
\usepackage{tikz}
\usepackage{setspace}
\usepackage{cancel}
\usepackage{array}
\usepackage{tabulary}
\usepackage{doi}
\definecolor{purple}{rgb}{1,0,1}
\definecolor{lime}{HTML}{A6CE39} 

\newcommand{\blue}[1]{{\color{blue} #1}}

\definecolor{lime}{HTML}{A6CE39}
\newcommand{\orcidicon}{%
	\begin{tikzpicture}
	\draw[lime, fill=lime] (0,0) 
		circle [radius=0.16] 
		node[white] {{\fontfamily{qag}\selectfont \tiny ID}};
	\draw[white, fill=white] (-0.0625,0.095) 
		circle [radius=0.007];
	\end{tikzpicture}
	\hspace{-5mm}
}
\newcommand\orcidChris{{\href{https://orcid.org/0009-0000-3953-0461}{\orcidicon}}}
\newcommand\orcidMatt{{\href{https://orcid.org/0000-0003-1088-6485}{\orcidicon}}}

\newcommand{\be}{\begin{equation}}
\newcommand{\ee}{\end{equation}}

\def\sign{\mathrm{{sign}}}
\begin{document}
\newcommand{\arXiv}[1]{arXiv:\href{https://arxiv.org/abs/#1}{\color{blue}#1}}

\title{\vspace{-25pt}\huge{
\blue{The spacetime geodesy\\
 of perfect fluid spheres}
}}


\author{\Large Christopher Simmonds\!\orcidChris\;}
\emailAdd{chris0simmonds@gmail.com}
\author{\!\!and \Large Matt Visser\!\orcidMatt$^{\dagger}$}
\emailAdd{matt.visser@sms.vuw.ac.nz}
\affiliation{School of Mathematics and Statistics, Victoria University of Wellington, \\
\null\qquad PO Box 600, Wellington 6140, New Zealand.}
\affiliation{$^\dagger$ Corresponding author.}
\renewcommand{\arXiv}[1]{arXiv:\href{https://arxiv.org/abs/#1}{\color{blue}#1}}
\def\L{{\mathcal{L}}}

\abstract{ \\
Herein we  shall  argue for the utility of ``spacetime geodesy'', a point of view where one delays 
as long as possible worrying about dynamical equations, in favour of the maximal utilization 
of both symmetries and geometrical features. This closely parallels Weinberg's distinction 
between ``cosmography'' and ``cosmology'', wherein maximal utilization 
of both the symmetries and geometrical features of 
 Friedmann--Lema\^i{}tre--Robertson--Walker
(FLRW) spacetimes is emphasized. 
This ``spacetime geodesy'' point of view is particularly useful in those situations where, 
for one reason or another, the dynamical equations of motion are either uncertain or completely unknown. 
Several specific  examples are discussed---we shall illustrate what can be done by considering the physics implications of demanding spatially isotropic Ricci tensors as a way of automatically implementing the (isotropic) perfect fluid condition, without committing to a specific equation of state. We also consider the structure of the Weyl tensor in spherical symmetry, with and without  the (isotropic) perfect fluid condition, and relate this to the notion of ``complexity''.
In closing, we indicate some ways in which these considerations might be further generalized to more physically complicated (and technically very much more complicated) situations such as axisymmetric spacetimes.

\bigskip
\noindent
{\sc Date:} Monday 20 October 2025;  1 December 2025; \LaTeX-ed \today

 
 \bigskip
\noindent{\sc Published}:  Symmetry {\bf 17 \#12}  (2025) 2043; \doi{10.3390/sym17122043}.

\bigskip
\noindent

}

\maketitle
\def\tr{{\mathrm{tr}}}
\def\diag{{\mathrm{diag}}}
\def\cof{{\mathrm{cof}}}
\def\pdet{{\mathrm{pdet}}}
\def\QED{ {\hfill$\Box$\hspace{-25pt}  }}
\def\d{{\mathrm{d}}}
\def\sign{\hbox{sign}}

\parindent0pt
\parskip7pt

\clearpage
\null
\vspace{-75pt}
\section{Introduction}
\def\theta{\vartheta}
\def\phi{\varphi}
\vspace{-7pt}
When trying to extend and go beyond ``known physics'', should you use a dynamical model or a kinematical model?
That depends critically on just how much you actually know (or think you know) about the details of the system in question, {and how you might wish to modify it}.
Can you find an actual solution to a reasonably well defined set of equations of motion? 
If so, great!
Otherwise, the~next best option is to build a purely kinematic model, using symmetries and general principles. Though careful thought and discretion is highly advised, one should try to minimize the speculative aspects of the model. (A common aphorism in the experimental community is this: ``Do not adjust more than one knob on your equipment at a time''.  The~theoretical community would be well advised to adopt the related maxim: ``No more than one miracle at a time'').   
Extract as much information as possible without committing to a specific choice of dynamics. 
Sometimes, once you have {somehow developed} an interesting model,
you can even reverse-engineer a suitable Lagrangian.  (A potential drawback is that reverse engineering is often fine-tuned and fragile).

\subsection{Cosmography \emph{versus} cosmology }

A now fully mainstream example of this behaviour is Weinberg's distinction between ``cosmography'' and ``cosmology''~\cite{Weinberg:1972}. Cosmography was developed in an attempt to extract as much information as possible from {adopting} the geometry of { Friedmann--Lema\^i{}tre--Robertson--Walker
(FLRW)} spacetimes as a zeroth-order approximation to reality, while eschewing, (as far as possible), any and all arguments regarding the cosmological equation of state. See, for~instance, the~extensive discussion in references~\cite{Weinberg:1972, Bamba:2012, Capozziello:2019, Capozziello:2011, Capozziello:2008, Mandal:2020, Capozziello:2020, 
Visser:2004,  Shafieloo:2012, Sathyaprakash:2009, Aviles:2012a, Lavaux:2011, Dunsby:2015, Courtois:2013, Cattoen:2008, Capozziello:2008GRB, Vitagliano:2009, Cattoen:2007, Xia:2011, Aviles:2012b, Capozziello:2017ddd, Luongo:2011zz, Lobo:2020, Visser:2009, Heinesen:2021, Apostolopoulos:2016, Apostolopoulos:2024}. Some {of the} implementations of this general idea instead speak in terms of ``cosmokinetics'' versus ``cosmodynamics''~\cite{Blandford:2004, Nair:2011, Shapiro:2005, Linder:2008, Cattoen:2007xx, Visser:2003jerk}. Work along these lines has by now led to well over 350 published scientific articles.
Note that adopting a ``cosmographic'' point of view implies that one need not make any specific a priori 
commitment to any specific variant of cosmological inflation, nor, if~one wishes to consider any form of ``modified gravity'', need one make  specific a~priori commitment to any specific variant of Horava gravity, or~of any specific version of $f(R)$, 
 $f(R,T)$, $f(R,T,Q)$ gravity or the like. The~cosmographic framework is sufficiently flexible to allow one to deal with all of these situations in a theory-agnostic manner.

\subsection{Synge's G-method \emph{versus} Synge's S-method }
\vspace{-5pt}
Synge's G-method of ``solving'' the Einstein equations amounts to identifying some physically or pedagogically interesting spacetime, and~then using the geometry of that spacetime to calculate the Einstein tensor. Thereby, (assuming applicability of the Einstein equations), one can deduce the required stress-energy required to support that spacetime~\cite{Synge:1961}. When used with care and discretion, this process can lead to interesting results. 

When used without care or discretion the results can, however,  be  unfortunate.  See, for~instance, the~criticism and discussion in reference~\cite{Ellis:2023}. In~contrast, Synge's {S\nobreakdash-method} is much more traditional, and~is based on iterating approximate solutions of known equations of~motion.

Though not originally phrased in this particular manner, the~core of the extensive body of work on ``traversable wormholes''~\cite{Morris-Thorne, MTY, Visser:1989a, Visser:1989b, Hochberg:1990, Frolov:1990, Roman:1992, Hochberg:1992, Cramer:1994, Visser:1995,  Poisson:1995, Kar:1994, Kar:1995, 
Hochberg:1995, Hochberg:1997, Visser:1997, Hochberg:1998,  
Krasnikov:1999, Armendariz-Picon:2002, Lemos:2003, Visser:2003, Kar:2004, Lobo:2005, Sushkov:2005, Harko:2013, Damour:2007, Lobo:2007, Martin-Moruno:2009, Konoplya:2010, Nakajima:2012, Lobo:2017, Roman:2004, Boonserm:2018, DuttaRoy:2019, Kar:2022}, ``warp drives''~\cite{Alcubierre:1994, Ford:2000, Lobo:2004-warp, Everett:1995, Everett:1997, Pfenning:1997, Clark:1999, Natario:2001, Lobo:2002, Hiscock:1997, Finazzi:2009, Barcelo:2010-F, Santiago:2021-warp, Lobo:2004-linearized, Coutant:2011, Shoshany:2019, Shoshany:2023, Alcubierre:2017-basics, Barcelo:2022-cpc, Schuster:2022-adm, Liberati:2016-mess, Barcelo:2022-aero, Barcelo:2010-impossible, Finazzi:2010, Schuster:2023-frenemies, Clough:2024}, and~even ``tractor beams''~\cite{Santiago:2021-tractor,Visser:2021-tractor} can be viewed as (by and large successful) applications of Synge's G-method (see also~\cite{Tippett:2013, Hiscock:2002, Obousy:2008}). Work along these lines has by now led to well over 600 published scientific articles.
As long as one keeps the speculative aspects of the physics under tight control, then useful information can be extracted. We particularly wish to emphasize the need for careful ``sanity checking'' to make sure one remains compatible with observational reality.

\subsection{Immediate plan of action}

Deferring dynamical considerations, at least until after one has a tolerable grasp on the kinematics, and similarly 
deferring considerations regarding the equation of state, until after one has a tolerable grasp on the spacetime geometry, can often be an ``intermediate'' but nevertheless useful route (the ``low road'') to significant progress. Below  we shall re-assess and re-analyze the venerable century old problem of the general relativistic perfect fluid sphere, (often used as a first approximation to stellar structure)~\cite{Schwarzschild-interior}. We shall do so from a purely geometrical perspective,  focussing on the spacetime geodesy (rather than the TOV equation and the EOS). 

\subsection{Long Term~Goals}

In the longer term, one would certainly be interested in dealing with mathematically more complicated  and more physically realistic spacetimes. However,~in both cosmography and in spacetime geodesy, there is very definitely a crucial trade-off between generality and~symmetry. 

For instance, grossly inhomogeneous cosmologies have essentially no symmetries, and there is little that can be said from a cosmographic perspective. On~the other hand, rotating stars might plausibly still retain axial symmetry, and~so a spacetime geodesy of (horizonless) stationary fluid spheroids might still be practical and useful. We shall leave such considerations for the~future.

\bigskip
\hrule\hrule\hrule

\clearpage

\hrule\hrule\hrule
\section{Spacetime geodesy of perfect fluid spheres: Framework}

Terrestrial geodesy, or~terrestrial geodetics, is the science of measuring, studying,  and~representing the geometry, gravity, and~spatial orientation of planet Earth. It is called planetary geodesy when studying other astronomical bodies, such as planets or the like. By~extension we shall adopt the phrase ``spacetime geodesy'' to describe the process of probing the geometry, gravity, and~shape of some specified~spacetime.

A particularly popular spacetime to study, because~it is a zeroth order approximation to a stable star,  is the static spherically symmetric perfect fluid sphere. The~first explicit analysis of such an object was that of Schwarzschild's constant density star~\cite{Schwarzschild-interior}. First, using only the static and spherically symmetric conditions,  geometrically this means the spacetime has a line element which in ``area coordinates'' can be put into the 
form 
\begin{equation}
ds^2 = - e^{-2\Phi(r)} dt^2 + {dr^2\over 1-2m(r)/r} + r^2 \; d\Omega^2.
\end{equation}
Area coordinates are often misattributed to Schwarzschild~\cite{Schwarzschild-interior, Schwarzschild-exterior}, though~they should more appropriately be attributed to Droste~\cite{Droste:1916,Droste:2002a,Droste:2002b} and Hilbert~\cite{Hilbert:1916}. Area coordinates have the useful property that the area of a sphere of ``coordinate radius'' $r$ is simply $A(r)=4\pi r^2$.
This line element, and~variations thereof, have been the inspiration for an enormous output of the literature. See, in particular, references~\cite{TOV, Delgaty:1998, Stephani:2003, MacCallum:2006, Griffiths:2009, Hawking-Ellis, Finch-Skea, Tolman:1939, Barraco:2002, Mak:2013, Lake:2008} and, more generally, references~\cite{Rahman:2001, Martin:2003, Boonserm:2005, Boonserm:2005-msc, Boonserm:2006, Boonserm:2007, Boonserm:2007b, Kinreewong:2016, Boonserm:2019-siri, Boonserm:2021, Mantica:2024}.

We shall now seek to abstract the purely geometrical properties of these various and sundry models with a view to developing an appropriate notion of spacetime geodesy. 
More specifically, and~more abstractly, writing the coordinates as $x^a = (t,x^i)$,  we quite generally have
\begin{equation}
ds^2 = - e^{-2\Phi(r)} dt^2 + g_{ij}(x^k)\;dx^i \; dx^j.
\end{equation}
Then, for the metric,
\begin{equation}
g_{ab} = \left[\begin{array}{c|c}- e^{-2\Phi(r)} & 0 \\ \hline 0 & g_{ij}(x^k)\end{array}\right].
\end{equation}
This is enough to guarantee that the only non-zero parts of the Riemann tensor are
$R_{titj}$ and  $R_{ijkl}$. In~turn, this implies block-diagonalization of the Ricci tensor:
\begin{equation}
R_{ab} = \left[\begin{array}{c|c} R_{tt} & 0 \\ \hline 0 & R_{ij}\end{array}\right].
\end{equation}

\subsection{Generalized eigenvalues}

Useful geometrical quantities are the coordinate independent generalized eigenvalues of the Ricci tensor defined by
\begin{equation}
\Lambda = \left\{ \lambda: \det\left( R_{ab} - \lambda \; g_{ab}\right) = 0 \right\}.
\end{equation}
Static spherical symmetry is enough to guarantee $\Lambda = \{\lambda_0,\lambda_1, \lambda_2,\lambda_2\}$. The geometrical equivalent of the pressure isotropy condition inherent in demanding a perfect fluid sphere is a condition demanding spatial isotropy of the Ricci tensor --- that all three spatial eigenvalues be equal: $\Lambda = \{\lambda_0,\lambda_1, \lambda_1,\lambda_1\}$. This in turn implies
\begin{equation}
R_{ij} = \lambda_1 \; g_{ij} = {1\over3} (g^{kl} R_{kl}) \; g_{ij}.
\end{equation}
This purely geometrical statement now encodes the essential physics of a  perfect fluid sphere, without having to resort to detailled investigation of what Einstein referred to as ``base matter''. Everything is now encoded in the ``marble'' of spacetime geometry. 
For practical computations it is often preferable to work with the ordinary eigenvalue problem
\begin{equation}
\Lambda = \left\{ \lambda: \det\left( R^a{}_b - \lambda \; \delta^a{}_b\right) = 0 \right\}.
\end{equation}
The only potential disadvantage of the ``mixed-index'' $T^1_{\;1}$ formulation is that in situations more general than those currently under consideration there is a risk that $R^a{}_b$, because it no longer need be symmetric, might lead to non-trivial Jordan normal forms~\cite{Martin-Moruno:2017-Rainich,Martin-Moruno:2018-core}. Fortunately this is not an issue in static spherical symmetry.

In this ``mixed-index'' $T^1_{\;1}$ formulation the spatial Ricci isotropy condition becomes
\begin{equation}
R^i{}_j = \lambda_1 \; \delta^i{}_j = {1\over3} (R^k{}_k) \; \delta^i{}_j.
\end{equation}
This now completely specifies the purely geometrical spacetime geodesy framework we will more fully investigate below. 

\subsection{Physical interpretation of  the ``mixed-index'' $T^1_{\;1}$ and  $T^2_{\;2}$ components}

It is also worthwhile to emphasize that, for any diagonal metric, provided the Ricci and Einstein tensors are similarly diagonal,
adopting the ``mixed-index'' $T^1_{\;1}$ components for the Ricci $R^a{}_b$ or Einstein $G^a{}_b$ tensors is effectively equivalent to adopting an orthonormal tetrad basis, but without the hassle of actually defining the orthonormal tetrad.
To justify this, note that when everything is diagonal
\begin{equation}
R_{ab} = e_a{}^{\hat a} \; R_{\hat a \hat b} \; e_b{}^{\hat b} = 
 \diag\{ R_{\hat a\hat a} \;  e_a{}^{\hat a}  \; e_a{}^{\hat a}  \} 
 = \diag\{ R_{\hat a\hat a}  \; |g_{aa}|  \}.
\end{equation}
Equivalently, when everything is diagonal
\begin{equation}
R^a{}_{b} = e^a{}_{\hat a} \; R^{\hat a}{}_{\hat b} \; e_b{}^{\hat b} = 
 \diag\{ R^{\hat a}{}_{\hat a} \;  e^a{}_{\hat a}   \; e_a{}^{\hat a}  \} 
 = \diag\{ R^{\hat a}{}_{\hat a} \} = R^{\hat a}{}_{\hat b}.
\end{equation}
That is, under these circumstances, we have $R^a{}_{b} = R^{\hat a}{}_{\hat b}$.
 (For this reason, older textbooks sometimes refer to $R^a{}_b$ and $G^a{}_b$ as the ``physical'' components of the Ricci and Einstein tensors.) 

This trick also extends to the Riemann and Weyl tensors, provided the  Riemann and Weyl tensors are diagonal under the $6\times6$ Petrov classification, where
 the ``mixed-index'' $T^2_{\;2}$ components $R^{ab}{}_{cd}$ and $C^{ab}{}_{cd}$ now enjoy all of the benefits of adopting an orthonormal tetrad basis, but without the hassle.
To see how this works view the antisymmetric combination $[ab]$ or $[cd]$ as a compound index on six-dimensional space according to the recepie
\begin{equation}
1 \leftrightarrow [tr]; \quad 2 \leftrightarrow [t\theta]; \quad 
3 \leftrightarrow [t\phi]; \quad  4 \leftrightarrow [r\theta]; \quad 
5 \leftrightarrow [r\phi]; \quad 6 \leftrightarrow [\theta\phi].
\end{equation}
Then the identification $R_{abcd}\longleftrightarrow R_{AB}$ and $C_{abcd}\longleftrightarrow C_{AB}$ allows you to reinterpret the Riemann and Weyl tensors as $6\times6$  symmetric matrices. If these $6\times6$  matrices (and the metric) are diagonal then for instance 
\begin{equation}
R^{ab}{}_{cd} = 
e^a{}_{\hat a} \; e^b{}_{\hat b} \;  R^{\hat a\hat b}{}_{\hat c\hat d} 
\; e_c{}^{\hat c} \; e_d{}^{\hat d} 
= 
 \diag\{ R^{\hat a\hat b}{}_{\hat a\hat b } \;  (e^a{}_{\hat a} e^b{}_{\hat b} )  
 \; (e_c{}^{\hat a} \; e_d{}^{\hat b} )  \} 
 = \diag\{ R^{\hat a\hat b}{}_{\hat a\hat b} \} 
 = R^{\hat a\hat b}{}_{\hat c\hat d}.
\end{equation}
That is, under these circumstances, $R^{ab}{}_{cd} = R^{\hat a\hat b}{}_{\hat c\hat d}$, and likewise $C^{ab}{}_{cd} = C^{\hat a\hat b}{}_{\hat c\hat d}$.

Unfortunately, this trick is not truly fundamental --- it certainly fails whenever the metric is non-diagonal, (e.g. Kerr, Kerr--Newman, and more generically in rotating spacetimes), and can also be problematic in time dependent situations where both the Einstein and Ricci tensors acquire off-diagonal flux components, and the Riemann and Weyl tensors need no longer be Petrov diagonal. The  ``mixed-index'' $T^1_{\;1}$ and $T^2_{\;2}$ still exist --- but they no longer have any ``nice''  straightforward interpretation in terms of  an orthonormal tetrad basis.
Still, as long as one is aware of the limitations thereof, this trick is a useful way of simplifying explicit calculations. 

\subsection{Pragmatics}

At a purely pragmatic level we can now proceed simply by writing down the  line element for a static spherically symmetric spacetime and solving the spatial isotropy condition for the Ricci tensor:
\begin{equation}
R^r{}_r = R^\theta{}_\theta = R^\phi{}_\phi. 
\end{equation}
Of course spatial isotropy for the Ricci tensor is completely equivalent to spatial isotropy for the Einstein tensor
\begin{equation}
G^r{}_r = G^\theta{}_\theta = G^\phi{}_\phi. 
\end{equation}
This spatial isotropy condition supplies one ordinary differential equation relating the metric components. 
Depending on the precise choice of coordinates, and ways of representing the metric components,  this  ordinary differential equation might be explicitly solvable. 

(It should be noted that  even symbolic calculation systems such as {\sf Maple} or {\sf Mathematica} will often require significant human intervention to obtain a reasonably tractable result.) We shall present a number of explicit examples below.

\bigskip
\hrule\hrule\hrule



\section{Spacetime geodesy of perfect fluid spheres: Examples}

Below we present several specific examples of spacetime geodesy --- of various levels of generality and practicality. The primary goal will be to generate purely geometrical models of perfect fluid spheres, focussing on the spacetime geometry, and (temporarily) delaying dynamical considerations. 

\subsection{Example 1: Two free functions --- purely integral version}
We start with a rather non-trivial (certainly not \emph{a priori} obvious) example of spacetime geodesy.
Take two  arbitrary integrable functions, $f(r)$ and $w(r)$, plus an arbitrary constant $K_0$, and  consider this line-element:
\begin{eqnarray}
\hspace{-7pt}
ds^2 &=& - \exp\left(-2\int f(r) dr\right)dt^2 
\nonumber\\
&&+\left\{ {
 f(r)^2 [ 1+w(r)]^2 \exp\left( - 2 \int f(r){[1-w(r)]\over [ 1+w(r)]} dr\right)
\over
2\int  f(r) [ 1+w(r)] 
\exp\left( - 2 \int{ f(r) [1-2w(r)-w(r)^2]\over [ 1+w(r)]} dr\right) dr - K_0}\right\} dr^2 \qquad
\nonumber\\
&&
+ \exp\left(-2\int  f(r) w(r)dr\right) [ d\theta^2+\sin^2\theta \; d\phi^2].
\label{E:master}
\end{eqnarray}

This always has a spatially isotropic Ricci tensor satisfying $R^r{}_r = R^\theta{}_\theta = R^\phi{}_\phi$ (and therefore, assuming the usual Einstein equations, is suitable for describing a perfect fluid sphere, and even if one does not wish to assume the usual Einstein equations, this is nevertheless a geometrically interesting spacetime with locally isotropic spatial curvature).

Verifying this is relatively easy: Just feed the line element into {\sf Maple} and check that $R^r{}_r = R^\theta{}_\theta = R^\phi{}_\phi$. (Some human intervention is still required.)  On the other hand, \emph{finding} this form of the line element is somewhat tedious and requires some inspired guesswork and more than a little direct human intervention. 
Let us see how to derive this result. 

\enlargethispage{20pt}
 Let us start with the metric in its most general form
\begin{equation}
ds^2 = g_{tt}(r) dt^2 + g_{rr}(r) dr^2 + g_{\theta\theta}(r)\, [ d\theta^2+\sin^2\theta \; d\phi^2].
\end{equation}
This respects spherical symmetry and time independence but at this stage imposes no additional  constraint.

Before we do anything else let us consider the null convergence condition (NCC). \\
In complete generality
\begin{equation}
R^t{}_t - R^r{}_r = - {g_{\theta\theta}''\over  g_{\theta\theta}\; g_{rr}}
+{[g_{\theta\theta}']^2\over  2g_{\theta\theta}^2 \; g_{rr}}
+{g_{\theta\theta}' \over 2 g_{rr} \; g_{\theta\theta} } 
\left[ {g_{tt}'\over g_{tt}} + {g_{rr}'\over g_{rr}} \right].
\end{equation}
Thence at any local extremum of $g_{\theta\theta}$ we have
\begin{equation}
R^t{}_t - R^r{}_r \to - {g_{\theta\theta}''\over  g_{\theta\theta} g_{rr}}.
\end{equation}
This implies that local minima of $g_{\theta\theta}$ violate the NCC (implying, in standard Einstein gravity, violation of the null energy condition, NEC). Violations of the NCC are not entirely forbidden by quantum physics,  but are certainly a significant departure from ``usual physics''~\cite{Curiel:2014, Martin-Moruno:2017, Barcelo:2002, Borissova:2025a, Borissova:2025b}. 
These considerations are important if you either wish to implement (or prevent) formation of a traversable wormhole throat~\cite{Morris-Thorne, MTY}.

Now consider the ordinary differential equation (ODE)
\begin{equation}
R^r{}_r - R^\theta{}_\theta = 0.
\end{equation}
Viewed as an ODE for for $g_{rr}(r)$, this ODE is a Bernoulli ODE , so nominally solvable.
There is an art to choosing $g_{tt}(r)$, $g_{rr}(r)$, $g_{\theta\theta}(r)$ to simplify this as much as possible.
Let us now choose
\begin{equation}
ds^2 = g_{tt}(r) dt^2 + {dr^2\over B(r)} + g_{\theta\theta}(r)\, [ d\theta^2+\sin^2\theta \; d\phi^2].
\end{equation}
The ODE $R^r{}_r - R^\theta{}_\theta = 0$ is now a first order linear ODE for $B(r)$. 
Write this ODE in the form
\begin{equation}
K_1(r) B'(r) + K_2 B(r) + K_3(r)=0.
\end{equation}
Then one finds
\begin{eqnarray}
K_1(r) &=& {1\over4} {d \ln[g_{tt}(r) g_{\theta\theta}(r)] \over dr};
\\
K_2(r) &=&  {1\over2}  {d^2 \ln[g_{tt}(r) g_{\theta\theta}(r)] \over dr^2} 
+{1\over4}  {d \ln[g_{tt}(r)] \over dr} \;  {d \ln[g_{tt}(r)/ g_{\theta\theta}(r)] \over dr} ;
\\
K_3(r) &=&{1\over g_{\theta\theta}(r)}.
\end{eqnarray}
Looking at $K_1(r)$ and $K_2(r)$ suggests that it might be useful to set
\begin{equation}
g_{tt}(r) = - \exp\left(-2\int f(r) dr\right); \qquad
g_{\theta\theta}(r) =  \exp\left(-2\int h(r) dr\right);
\end{equation}
for arbitrary smooth functions $f(r)$ and $h(r)$.
So the line element becomes
\begin{equation}
ds^2 = - \exp\left(-2\int f(r) dr\right)dt^2 +{dr^2\over B(r)}
+ \exp\left(-2\int h(r) dr\right) [ d\theta^2+\sin^2\theta \; d\phi^2].
\end{equation}

The ODE imposing spatial isotropy, $R^r{}_r - R^\theta{}_\theta = 0$, when viewed as an ODE for the function $B(r)$, and written as $K_1(r) B'(r) + K_2 B(r) + K_3(r)=0$, now yields
\begin{eqnarray}
K_1(r) &=& -{1\over2} [ f(r)+h(r)];
\\
K_2(r) &=& -[f'(r)+h'(r)] +f(r)[f(r)-h(r)];
\\
K_3(r) &=& \exp\left(+2\int h(r) dr\right).
\end{eqnarray}
We know that this ODE
has the explicit solution
\begin{equation}
B(r) = -\left( \int {K_3(r)\over K_1(r)} \;
\exp\left(\int{K_2(r)\over K_1(r)} dr \right)  dr + K_0 \right)
\exp\left(-\int{K_2(r)\over K_1(r)} dr \right).
\end{equation}

Note the following:
\begin{eqnarray}
\int{K_2(r)\over K_1(r)} dr &=&
\int{ [f'(r)+h'(r)] -f(r)[f(r)-h(r)]\over {1\over2} [ f(r)+h(r)]} dr
\\
&=& 2 \ln[ f(r)+h(r)] - 2 \int{ f(r)[f(r)-h(r)]\over [ f(r)+h(r)]} dr.
\end{eqnarray}
Thence
\begin{equation}
\exp\left(\int{K_2(r)\over K_1(r)} dr \right)
=
[ f(r)+h(r)]^2 \exp\left( - 2 \int{ f(r)[f(r)-h(r)]\over [ f(r)+h(r)]} dr\right).
\end{equation}
Furthermore 
\begin{eqnarray}
&&{K_3(r)\over K_1(r)}\exp\left(\int{K_2(r)\over K_1(r)} dr \right)
\nonumber\\
&&\qquad\qquad=-2  \exp\left(+2\int h(r) dr\right)
 [ f(r)+h(r)] \exp\left( - 2 \int{ f(r)[f(r)-h(r)]\over [ f(r)+h(r)]} dr\right)
\nonumber\\
&&\qquad\qquad = 
-2 [ f(r)+h(r)] \exp\left( - 2 \int{ [f(r)^2-2f(r)h(r)-h(r)^2]\over [ f(r)+h(r)]} dr\right).
\end{eqnarray}
Thence
\begin{equation}
g_{rr}(r) = {1\over B(r)} =
{
[ f(r)+h(r)]^2 \exp\left( - 2 \int{ f(r)[f(r)-h(r)]\over [ f(r)+h(r)]} dr\right)
\over
2\int  [ f(r)+h(r)] 
\exp\left( - 2 \int{ [f(r)^2-2f(r)h(r)-h(r)^2]\over [ f(r)+h(r)]} dr\right) dr - K_0}.
\end{equation}

Therefore, we have the line element
\begin{eqnarray}
ds^2 &=& - \exp\left(-2\int f(r) dr\right)dt^2 
\nonumber\\ &&
+\left\{ {
[ f(r)+h(r)]^2 \exp\left( - 2 \int{ f(r)[f(r)-h(r)]\over [ f(r)+h(r)]} dr\right)
\over
2\int  [ f(r)+h(r)] 
\exp\left( - 2 \int{ [f(r)^2-2f(r)h(r)-h(r)^2]\over [ f(r)+h(r)]} dr\right) dr - K_0}\right\} dr^2
\nonumber\\ &&
+ \exp\left(-2\int h(r) dr\right) [ d\theta^2+\sin^2\theta \; d\phi^2] .
\label{E:slave}
\end{eqnarray}
\enlargethispage{40pt}
Finally, define $w(r)$ by setting $h(r)= f(r)w(r)$, to obtain the result previously announced in equation (\ref{E:master}). 
The good news is that we have a fully explicit line element always guaranteed to generate a spatially isotropic Ricci tensor (ditto a spatially isotropic Einstein tensor).  The not so good news is that considerable brute force was required to get to this result, and that the resulting curvature tensors are extremely clumsy to deal with.
Still, with 2 free functions at one's disposal, either $\{f(r),w(r)\}$ in equation (\ref{E:master}) or $\{f(r),h(r)\}$ in equation (\ref{E:slave}), one has very powerful algorithm for generating bespoke perfect fluid spheres.

\subsection{Example 2: Two free functions --- Integro-Differential version}
A perhaps slightly more tractable version of the above is to simply set
\begin{equation}
F(r) = \int f(r) dr; \qquad \qquad H(r) = \int h(r) dr,
\end{equation}
 so that the line element becomes
 \begin{equation}
ds^2 = - \exp\left(-2F(r)\right)dt^2 +{dr^2\over B(r)}
+ \exp\left(-2H(r) \right) [ d\theta^2+\sin^2\theta \; d\phi^2].
\end{equation}
Then, working directly from equation (\ref{E:slave}) we get
\begin{eqnarray}
ds^2 &=& - \exp\left(-2F(r)\right)dt^2 
\nonumber\\ &&
+\left\{ {
[ F'(r)+H'(r)]^2 \exp\left( - 2 \int F'(r){[F'(r)-H'(r)]\over [ F'(r)+H'(r)]} dr\right)
\over
2\int  [ F'(r)+H'(r)] 
\exp\left(2 H(r) - 2 \int F'(r){[F'(r)-H'(r)]\over [ F'(r)+H'(r)]}dr\right) dr - K_0}\right\} dr^2
\nonumber\\ &&
+ \exp\left(-2H(r) \right) [ d\theta^2+\sin^2\theta \; d\phi^2] .
\label{E:slave2}
\end{eqnarray}
This is again  has a spatially isotropic Ricci tensor satisfying $R^r{}_r = R^\theta{}_\theta = R^\phi{}_\phi$. 
Verifying this is now somewhat easier ---  {\sf Maple} requires significantly less human intervention.
Unfortunately the resulting curvature tensors are still extremely clumsy to deal with.
On the other hand, we still have two free functions available, so we still have room for using coordinate freedom to further simplify the situation.

Still, with 2 free functions at one's disposal,  $\{F(r),H(r)\}$ in equation (\ref{E:slave2}) , one again has very powerful algorithm for generating bespoke perfect fluid spheres.

\subsection{Example 3: One free function --- area coordinates}

The Hilbert--Droste area coordinates~\cite{Droste:1916, Droste:2002a, Droste:2002b, Hilbert:1916} amount to setting $g_{\theta\theta}=r^2$ so that the area of sphere of coordinate radius $r$ is simply $A(r)=4\pi \; r^2$.  There is a mild technical constraint involved here --- if $A(r)$ is not monotone increasing as one moves outwards then one needs multiple coordinate patches of this type, which overlap at wormhole throats or anti-throats. (In particular, as already noted, local minima of $g_{rr}(r)$, and hence local minima of $A(r)=4\pi \;g_{\theta\theta}(r)=4\pi \;r^2$ correspond to violations of the null convergence condition (NCC). Consequently, globally enforcing the existence of a single patch of area coordinates automatically precludes the existence of traversable wormholes.) 

In normal situations this is not an issue and one globally asserts the following.
 \begin{equation}
ds^2 = - \exp\left(-2F(r)\right)dt^2 +{dr^2\over B(r)}
+ r^2 [ d\theta^2+\sin^2\theta \; d\phi^2].
\end{equation}
(If one does want to allow the possibility of traversable wormholes then area coordinates should be used with care and discretion --- with wormhole throats and anti-throats being located right on the edge of these coordinate patches, it is easy to get confused. See references~\cite{defective, Feng:2023, Simmonds:2025} for some examples of potential pitfalls.)

Then, working directly from equation (\ref{E:slave2}) we get
\begin{eqnarray}
ds^2 &=& - \exp\left(-2F(r)\right)dt^2 
\nonumber\\ &&
+\left\{ {
[ 1- rF'(r)]^2 r^{-2} \exp\left( - 2 \int F'(r){[1+r F'(r)]\over [ 1- rF'(r)]} dr\right)
\over
-2\int  [1-r F'(r)] r^{-3}
\exp\left(- 2 \int F'(r){[1+rF'(r)]\over [1-r F'(r)]}dr\right) dr + K_0}\right\} dr^2
\nonumber\\ &&
+ r^2[ d\theta^2+\sin^2\theta \; d\phi^2] .
\label{E:slave3}
\end{eqnarray}
This is again  has a spatially isotropic Ricci tensor satisfying $R^r{}_r = R^\theta{}_\theta = R^\phi{}_\phi$. 
Verifying this is now considerably easier ---  in this situation {\sf Maple} requires essentially no direct human intervention. This result is essentially a repackaging of the Boonserm--Visser--Weinfurtner algorithm~\cite{Boonserm:2005, Boonserm:2006}, (see also~\cite{Boonserm:2005-msc, Boonserm:2007, Boonserm:2007b}),  which we now see is a special case of a considerably more general analysis.

\subsection{Example 4:  One free function --- redshift coordinates}

Redshift coordinates are somewhat unusual. First re-label $r \to z$, that is just a name change, but now additionally use coordinate freedom to choose $g_{tt}(z) = -1/(1+z)^2$ so that
\begin{equation}
ds^2 = - {dt^2\over(1+z)^2} + 
{dz^2\over B(z)} + \exp(-2 H(z))\, [ d\theta^2+\sin^2\theta \; d\phi^2]. 
\end{equation}
The proper time measured by a clock at position $z$ is then $\Delta\tau = {\Delta t\over 1+z}$, which means that signals emitted from the location $z$ will be redshifted by exactly an amount $z$ by the time they reach spatial infinity. ($T_{emitted}= T_\infty/(1+z)$.) 
The range of the $z$ coordinate will be $z\in[0,z_0]$ where $z_0$ is the ``central redshift'' of a signal emitted from the centre of the spacetime. Coordinates of this type do require the mild technical restriction that $g_{tt}(z)$ be a monotone as one moves outwards, corresponding to the gravitational field always being attractive. 

Then, working directly from equation (\ref{E:slave2}) we now get
\begin{eqnarray}
ds^2 &=& - (1+z)^{-2}\; dt^2 
\nonumber\\ &&
+\left\{ {
[ 1- (1+z)H'(z)]^2 [1+z]^{-2} \exp\left( +2 \int {[1+ (1+z)H'(z)]\over [ 1- (1+z)H'(z)]} {dz\over1+z}\right)
\over
-2\int  [1-(1+z) H'(z)] 
\exp\left(2H(z) + 2 \int{[1+ (1+z)H'(z)]\over [ 1- (1+z)H'(z)]}{dz\over1+z}\right) {dz\over1+z} + K_0}\right\} dz^2
\nonumber\\ &&
+ \exp(-2H(z))\; [ d\theta^2+\sin^2\theta \; d\phi^2] 
\label{E:slave4}
\end{eqnarray}
This is again  has a spatially isotropic Ricci tensor satisfying $R^r{}_r = R^\theta{}_\theta = R^\phi{}_\phi$. 
Verifying this is again relatively easy ---  {\sf Maple} requires essentially no direct human intervention.

\subsection{Example 5:  One free function --- exponential redshift coordinates}
It is also worthwhile to consider an ``exponential'' version of redshift coordinates with $g_{tt}(z) = - e^{-2z}$ and 
$T_{emitted}= e^{-z} \; T_\infty$.)
This is not precisely the standard definition of redshift, but ultimately it carries the same physical information and is somewhat easier to deal with mathematically.
We start from
\[
ds^2 = - e^{-2z} \;dt^2 + 
{dz^2\over B(z)} + \exp(-2 H(z))\, [ d\theta^2+\sin^2\theta \; d\phi^2].
\]
Then, working directly from equation (\ref{E:slave2}) we now get
\begin{eqnarray}
ds^2 &=& - e^{-2z}\; dt^2 
\nonumber\\ &&
+\left\{ {
[ 1+H'(z)]^2\exp\left( -2 \int {[1-H'(z)]\over [ 1+H'(z)]} {dz}\right)
\over
2\int  [1+H'(z)] 
\exp\left(2H(z) - 2 \int{[1-H'(z)]\over [ 1+H'(z)]}{dz}\right) {dz} + K_0}\right\} dz^2
\nonumber\\ &&
+ \exp(-2H(z))\; [ d\theta^2+\sin^2\theta \; d\phi^2] 
\label{E:slave5}
\end{eqnarray}
This is again  has a spatially isotropic Ricci tensor satisfying $R^r{}_r = R^\theta{}_\theta = R^\phi{}_\phi$. 
Verifying this is again relatively easy ---  {\sf Maple} requires essentially no direct human intervention.

\subsection{Discussion of these Perfect Fluid Spheres}

In this section, we have developed a general algorithm guaranteed to always generate spacetime geometries describing perfect fluid spheres---this would only be a first step in investigating the physical properties of such spacetimes~\cite{Delgaty:1998}. The~isotropy condition, while central to the effort, is only part of what makes perfect fluid spheres physically acceptable~\cite{Delgaty:1998, Stephani:2003, MacCallum:2006, Griffiths:2009, Hawking-Ellis, Finch-Skea,  Tolman:1939, Barraco:2002, Mak:2013, Lake:2008}. Overall, this spacetime geodesy approach is a very practical and pragmatic first step in any such~analysis. 

\bigskip
\hrule\hrule\hrule

\clearpage

\bigskip
\hrule\hrule\hrule

\section{Weyl tensor and Weyl scalar in spherical symmetry}

A quite remarkable feature of  four-dimensional spacetime geodesy is that for spherical symmetry the Weyl tensor is particularly  and perhaps unexpectedly simple.

\subsection{Generalities}
Even allowing for the possibility of time dependence in the spacetime geometry, the Weyl tensor in four-dimensional spherical symmetry always takes the very simple form 
\begin{equation}
C^{ab}{}_{cd} = Z^{ab}{}_{cd} \;\; C_0(r,t),
\end{equation}
where the $T^2_{\;2}$  tensor $Z^{ab}{}_{cd}$ has only fixed integer-valued components, whereas  the common scalar $C_0(r,t)$ depends on the specific details of the spacetime. (And $C_0(r,t)$ will soon be seen to be  particularly simple for static fluid spheres.) 

Furthermore, since the single quantity $C_0(r,t)$ completely characterizes the deviation from conformal flatness, it also completely characterizes the lensing properties of any spherically symmetric spacetime. 

Specifically, it is convenient for this section to set
\begin{equation}
ds^2 = - \exp(-2\Phi(r,t))\left(1-{2m(r,t)\over r}\right) dt^2 + {dr^2\over 1-2m(r,t)/r} 
+ r^2 [d\theta^2 +\sin^2\theta\; d\phi^2].
\end{equation}
Now explicitly calculate the $T^2_{\;2}$  Weyl tensor $C^{ab}{}_{cd} $ and (by inspection) note that it is always of the form  $C^{ab}{}_{cd} = Z^{ab}{}_{cd} \;\; C_0(r,t)$, with
\begin{equation}
Z^{tr}{}_{tr} = Z^{\theta\phi}{}_{\theta\phi} = 2;
\end{equation}
\begin{equation}
Z^{t\theta}{}_{t\theta} = 
Z^{t\phi}{}_{t\phi} = 
Z^{r\theta}{}_{r\theta}  =
Z^{r\phi}{}_{r\phi} =  -1;
\end{equation}
and
\begin{equation}
C_0(r,t) = {1\over6}\left(G^t{}_t + G^r{}_r - G^\theta{}_\theta\right) + {m(r,t)\over r^3}.
\end{equation}
(All other components, not related by the standard Weyl symmetries, vanish.)
Note that this simple formula for $C_0(r,t)$ holds even though the time-dependent Einstein tensor $G^a{}_b$ has two non-zero off-diagonal components $G^t{}_r$ and $G^r{}_t$. 
Note that whereas the Einstein tensor components $G^t{}_t$, $G^r{}_r$, and $G^\theta{}_\theta$, by construction depend only on local physics, the quantity $C_0(r,t)$ also explicitly depends on one quasi-local concept --- the Misner--Sharp quasi-local mass $m(r,t)$. 

A side-effect of these results is that  in four-dimensional spherical symmetry the Weyl scalar is always a perfect square
\begin{equation}
W = C_{abcd} \; C^{abcd} = [Z_{abcd} \; Z^{abcd} ] \; [C_0(r,t)]^2 = 48 \;[C_0(r,t)]^2. 
\end{equation}

For the dual Weyl tensor $^*C^{ab}{}_{cd} = {1\over2} \epsilon^{ab}{}_{ef} \; C^{ef}{}_{cd} =  
{}^*\!Z^{ab}{}_{cd}\; C_0(r,t)$ we have
\begin{equation}
^*Z^{tr}{}_{\theta\phi}=  2; \qquad ^*Z^{r\theta}{}_{\phi r}=  {}^*Z^{t\phi}{}_{r\theta}=  -1;
\end{equation}
\leftline{(all other components, not related by the standard dual-Weyl symmetries, vanish).}

Defining $V$ to be a unit vector in the $t$ direction,  the ``electric'' part of the 
Weyl tensor is defined to be $E^a{}_b = C^{ac}{}_{bd} \; V_c V^d$.
Consequently $E^a{}_b =  [Z_E]^a{}_b \; C_0$,  where
the only non-zero components of $[Z_E]^a{}_b$ are
\begin{equation}
[Z_E]^r{}_r = -2; \qquad [Z_E]^\theta{}_\theta = [Z_E]^\phi{}_\phi = 1.
\end{equation}
Thence
\begin{equation}
[Z_E]^a{}_b = \left[\begin{array}{c|ccc}
0 & 0 & 0 & 0\\
\hline
0 & -2 & 0 & 0\\
0 & 0 & +1 & 0\\
0 & 0 & 0 & +1\\
\end{array}\right].
\end{equation}

(Note the tensor $[Z_E]^a{}_b$ is traceless.)
In contrast, it is easy to see that in spherical symmetry the ``magnetic'' part of the Weyl tensor $H^a{}_b = {}^*C^{ac}{}_{bd} \; V_c V^d$ vanishes identically. All in all, spherical symmetry very tightly constrains the Weyl tensor. 

We could also work with the $6\times 6$ Petrov embedding of $Z^{ab}{}_{cd} \longleftrightarrow Z^A{}_B$ in which case
\begin{equation}
Z^A{}_B = \left[\begin{array}{c|cc|cc|c} 
+2&0&0&0&0&0\\
\hline
0&-1&0&0&0&0\\
0&0&-1&0&0&0\\
\hline
0&0&0&-1&0&0\\
0&0&0&0&-1&0\\
\hline
0&0&0&0&0&+2\\
\end{array}\right]
\end{equation}

Finally, another way of characterizing the tensor structure is this: 
Ordering the coordinates as $\{t,r,\theta,\phi\}$ define the $T^2_{\;2}$ tensor:
\begin{equation}
\Delta^a{}_b= \left[\begin{array}{rr|rr}  
+1&0&0& 0 \\ 0&+1&0&0 \\ \hline 0&0&-1& 0 \\  0&0&0& -1 \\ 
\end{array} \right].
\end{equation}
This tensor squares to the ordinary Kronecker delta, $\Delta^a{}_b \Delta^b{}_c = \delta^a{}_c$, but its trace is zero $\Delta^a{}_a=0$. 
Then it is a easy exercise to check
\begin{equation}
Z^{ab}{}_{cd}= 
{3\over2} ( \Delta^a{}_c \Delta^b{}_d  - \Delta^a{}_d  \Delta^b{}_c )
+
{1\over2} ( \delta^a{}_c \delta^b{}_d  - \delta^a{}_d  \delta^b{}_c ).
\end{equation}
First check that this reproduces 
\begin{equation}
Z^{tr}{}_{tr} = Z^{\theta\phi}{}_{\theta\phi} = 2;
\end{equation}
and then check that this also reproduces
\begin{equation}
Z^{t\theta}{}_{t\theta} = 
Z^{t\phi}{}_{t\phi} = 
Z^{r\theta}{}_{r\theta}  =
Z^{r\phi}{}_{r\phi} =  -1.
\end{equation}
As a final consistency check note that when one takes the trace
\begin{equation}
Z^{ab}{}_{cb}=  -{3\over2} \delta^a{}_c +{1\over2} ([4-1] \delta^a{}_c) = 0.
\end{equation}
So all the Weyl symmetries, and the trace-free properties, are respected.

All in all, in spherical symmetry in (3+1) dimensions we have
\begin{equation}
C^{\mu\nu}{}_{\alpha\beta}= 
\left\{
{3\over2} ( \Delta^\mu{}_\alpha \Delta^\nu{}_\beta  - \Delta^\mu{}_\beta  \Delta^\nu{}_\alpha )
+
{1\over2} ( \delta^\mu{}_\alpha \delta^\nu{}_\beta  - \delta^\mu{}_\beta  \delta^\nu{}_\alpha )
\right\} C_0.
\end{equation}

\subsection{Weyl tensor for static perfect fluid spheres}

If we now consider static spacetimes, then the off-diagonal part of the Einstein (and Ricci) tensors vanish. Furthermore if (with perfect fluids in mind) we demand spatial isotropy for the Einstein tensor 
\begin{equation}
G^r{}_r = G^\theta{}_\theta = G^\phi{}_\phi, 
\end{equation}
then the common factor $C_0(r,t)$ simplifies considerably:
\begin{equation}
C_0(r,t) \to {1\over6}\;G^t{}_t  + {m(r)\over r^3} 
= -{m'(r)\over 3 r^2} + {m(r)\over r^3} = -{r (m(r)/r^3)'\over3}.
\end{equation}
This is a very simple function of the Misner--Sharp quasi-local mass. 
(Indeed $m(r)/r^3$ can be thought of as the ``average density'' inside coordinate radius $r$.) It is notable that the metric function $\Phi(r)$ has completely dropped out of the analysis.
Furthermore, for a constant density perfect fluid sphere, (Schwarzschild's constant density star, Schwarzschild's ``interior'' solution~\cite{Schwarzschild-interior}), the quantity $C_0(t)$, and hence the Weyl tensor itself, vanishes identically. 

One could even reverse-engineer $m(r)$ to set up a chosen profile $C_0(r)$ for the Weyl tensor. 
Merely set
\begin{equation}
m(r) = {\left( m_0 + 3 \int C_0(r) r^5  dr\right)\over r^3}.
\end{equation}
Unfortunately for this choice of mass profile the perfect fluid condition $G^r{}_r = G^\theta{}_\theta$ turns into a quite  nasty  nonlinear ODE for $\Phi(r)$. 

\subsection{PFDM --- not a perfect fluid}
The most critical  thing one really needs to know about PFDM (so-called ``perfect fluid dark matter'') is that it is not a perfect fluid~\cite{Kiselev1, Kiselev2}. Instead PFDM is (at best) interpretable as an anisotropic fluid --- indeed one should really call it a ``Kiselev fluid'',  and refer to KFDM The terminology originated in some (quickly corrected) awkward phrasing in reference~\cite{Kiselev0}, but some segments of the community continue to abuse terminology in this particular way. 

To really drive the message home, the eigenvalues  $\{ \lambda: \det(T^a{}_b -\lambda \delta^a{}_b)=0\}$ of the stress-enegy tensor for a perfect fluid are always of the form $\{\lambda_0,\lambda_1,\lambda_1,\lambda_1\}$, whereas the eigenvalues of Kiselev's stress-energy tensor are of the form $\{\lambda_0,\lambda_0,\lambda_1,\lambda_1\}$. These eigenvalue patterns are just not the same. (Except in the trivial case $\lambda_1=\lambda_0$, corresponding to ``vacuum energy''/cosmological constant $\{\lambda_0,\lambda_0,\lambda_0,\lambda_0\}$.)

The Kiselev geometry (at its most general) corresponds to the metric ansatz
\begin{equation}
ds^2 = -\left(1-{2m(r)\over r}\right) dt^2 + {dr^2\over 1-2m(r)/r} 
+ r^2 [d\theta^2 +\sin^2\theta\; d\phi^2],
\end{equation}
for which
\begin{equation}
G^t{}_t = G^r{}_r = -{2m'(r)\over r^2}; \qquad
G^\theta{}_\theta = G^\phi{}_\phi = -{m''(r)\over r}.
\end{equation}
Thence, for a Kiselev fluid
\begin{equation}
C_0(r) 
= {m''(r)\over 6 r} -{2m'(r)\over 3 r^2} + {m(r)\over r^3}
= {r^2 \,m''(r) - 4 r \, m'(r) +6m(r)\over 6 r^3}
= { r (m(r)/r^2)''\over 6},
\end{equation}
which generically is simply different from that of a perfect fluid.

\enlargethispage{20pt}
A much more specific form of Kiselev's spacetime is to take~\cite{Kiselev0}
\begin{equation}
m(r) = m_0\left( 1 + {k_{-1}\over r} - k_2 \, r^2 + {k_{-2w}\over r^{2w}}\right),
\end{equation}
with non-constant contributions to $m(r)$ coming from electrical charge, cosmological constant, and a specific version of the Kiselev fluid, respectively.

In this situation one obtains
\begin{equation}
C_0(r) = {m_0\over r^3} \left(1 + {2 k_{-1}\over r} +  {(1+w)(1+2w/3)k_{-2w}\over r^{2w}}\right).
\end{equation}
This  generically is simply different from what happens for a perfect fluid. (Except in the very special case $k_{-1}=0=k_{-2w}$, which corresponds to a pure cosmological constant.)

\subsection{Complexity factor}

The Weyl component $C_0(r)$ is almost identical to Herrara's ``complexity factor''. Indeed, assuming the usual Einstein equations, Herrara  defines $Y_{TF}$ as the equivalent \mbox{of~\cite{Herrera:2018a, Herrera:2018b, Herrera:2019a, Herrera:2019b, Herrera:2020, Herrera:2023, Herrera:2024, Herrera:2025, Herrera:2022, Herrera:2024b, Herrera:2024c, Herrera:2025b}}:
\begin{equation}
Y_{TF} = 4\pi (p_r-p_t)  + C_0(r).
\end{equation}
(See also various applications of related ideas in references~\cite{Casadio:2019, Sharif:2018, Yousaf:2020, Yousaf:2020b, Yousaf:2020c, Contreras:2021, Carrasco-Hidalgo:2021, Maurya:2022, Maurya:2022b, Arias:2021, Naseer:2024a, Sharif:2023, Nazar:2021, Kaur:2023, Naseer:2024b, Naseer:2024c, Khan:2024, Naseer:2025a, Naseer:2025b}.)

Certainly for any perfect fluid Herrara's complexity factor is identical to the Weyl component $C_0(r)$. Furthermore from a purely geometrical perspective
\begin{equation}
Y_{TF} = {1\over2} (G_r{}^r-G_\theta{}^\theta)  + C_0(r)
= {1\over6}G^t{}_t +  {2\over3}\left(G^r{}_r - G^\theta{}_\theta\right) + {m(r)\over r^3}.
\end{equation}
We note that both Herrara's ``complexity factor'' and the Weyl component $C_0(r,t)$ vanish for Schwarzschild's constant density perfect fluid star. 
Therefore,  both Herrara's ``complexity factor'' and the Weyl component $C_0(r,t)$ characterize deviations from Schwarzschild's constant density perfect fluid star. 
Furthermore,  note that any constant coefficient linear combination of $(G_r{}^r-G_\theta{}^\theta)$  and $C_0(r)$ also satisfies this property. 
Whether one prefers to use Herrara's ``complexity factor'' or  the Weyl component $C_0(r)$ seems largely a matter of taste --- though we would argue that the Weyl component $C_0(r)$ has a considerably clearer and direct geometrical interpretation in terms of the entire Weyl tensor vanishing identically. 

A more subtle approach would be to introduce a quadratic notion of complexity:
\begin{equation}
Q = \sqrt{ (G_r{}^r-G_\theta{}^\theta)^2  + C_0(r)^2}.
\end{equation}
This quadratic complexity vanishes \emph{if and only if} both $G_r{}^r-G_\theta{}^\theta$, (corresponding to a perfect fluid), and
then also $(m(r)/r^3)'=0$, (corresponding to constant density). So this quadratic complexity vanishes \emph{if and only if} one is dealing with Schwarzschild's constant density star, and we can interpret this quadratic complexity $Q$ as a ``distance'' between the geometry of interest and  Schwarzschild's constant density star. 

\subsection{Discussion of the~Weyl Tensor analysis}

In this section we have used a spacetime geodesy perspective to see that that in spherical symmetry the Weyl tensor takes on a very simple form --- effectively with only a single independent tensor component. We have seen that the Weyl tensor simplifies even further for perfect fluid spheres. Subsequently, we have considered the example of so-called PFDM, (perfect fluid dark matter), verifying that it is not a perfect fluid, and suggesting it be renamed KFDM, (Kiselev fluid dark matter).
Finally we have related these considerations to the notion of ``complexity''. 

\bigskip
\hrule\hrule\hrule

\section{Spacetime geodesy of anisotropic fluid spheres}

Finally, let us consider the spacetime geodesy of generic anisotropic fluid spheres \cite{Harko:2002,Tello-Ortiz:2020,Singh:2016}. In this situation, spacetime geodesy  is relatively uninteresting, because~once you allow for generic anisotropic fluids, in~principle, \emph{any} spherically symmetric spacetime is compatible with representing an anisotropic fluid sphere~\cite{Boonserm:mimic}. 

That is, for a generic anisotropic fluid sphere, because there is no longer any  geometrical constraint on the Einstein or Ricci tensors, the spacetime geodesy approach is not particularly useful. For some specific anisotropic fluid spheres, such as the Kiselev fluid considered above, one does have some geometrical constraint coming from an eigenvalue degeneracy in the Ricci tensor. But for a generic anisotropic fluid sphere, absent any such constraint, almost nothing can be said. 

\bigskip
\hrule\hrule\hrule

\section{Conclusions}

In this article we have advocated for the utility of a spacetime geodesy point of view.  
Delay, at least temporarily, concerns regarding equations of state and/or energy conditions,
until one at least has a clear geometrical picture of the spacetime of interest. 
We have illustrated the discussion with a number of examples, including both perfect fluid spheres and imperfect fluid spheres, and have placed rather remarkably tight constraints on the Weyl tensor in spherical symmetry.  

We again emphasize the theory-agnostic nature of the cosmographic/geodesy framework. Cosmography is indifferent to the particular implementation of cosmological inflation one wishes to impose, and~is indifferent to specific choices of ``modified gravity''. Cosmography provides a general framework for studying FLRW cosmologies without necessitating a specific choice of dynamics. Similarly, spacetime geodesy provides a general framework for studying the general features of localized clumps of gravitating matter without necessitating a specific choice of~dynamics.

As regards future plans, perhaps the most plausibly fruitful extension of these current ideas would be to developing a spacetime geodesy for the rotating axisymmetric spacetimes appropriate for modelling rotating stars. Given what we already know regarding rotating spacetimes, it is clear that such a project would be technically challenging, but~relatively straightforward, and~with high scientific impact. We hope to address such issues in the~future.


\bigskip
\bigskip
\hrule\hrule\hrule

\clearpage
\hrule\hrule\hrule
\addtocontents{toc}{\bigskip\hrule}

\vspace{-25pt}
\setcounter{secnumdepth}{0}
\section[\hspace{14pt}  References]{}
%


\begin{thebibliography}{99}

\bibitem{Weinberg:1972}
S.~Weinberg,
``Gravitation and Cosmology: Principles and Applications of the General Theory of Relativity'',
John Wiley and Sons, 1972,\\
ISBN 978-0-471-92567-5, 978-0-471-92567-5

\bibitem{Bamba:2012}
K.~Bamba, S.~Capozziello, S.~Nojiri and S.~D.~Odintsov,
``Dark energy cosmology: the equivalent description via different theoretical models and cosmography tests'',
Astrophys. Space Sci. \textbf{342} (2012), 155-228
\doi{10.1007/s10509-012-1181-8}
[\arXiv{1205.3421} [gr-qc]].

\bibitem{Capozziello:2019}
S.~Capozziello, R.~D'Agostino and O.~Luongo,
``Extended Gravity Cosmography'',
Int. J. Mod. Phys. D \textbf{28} (2019) no.10, 1930016
\doi{10.1142/S0218271819300167}
[\arXiv{1904.01427} [gr-qc]].

\bibitem{Capozziello:2011}
S.~Capozziello, V.~F.~Cardone, H.~Farajollahi and A.~Ravanpak,\\
``Cosmography in $f(T)$ gravity'',
Phys. Rev. D \textbf{84} (2011), 043527
\doi{10.1103/PhysRevD.84.043527}
[\arXiv{1108.2789} [astro-ph.CO]].

\bibitem{Capozziello:2008}
S.~Capozziello, V.~F.~Cardone and V.~Salzano,
``Cosmography of $f(R)$ gravity'',
Phys. Rev. D \textbf{78} (2008), 063504
\doi{10.1103/PhysRevD.78.063504}\\{}
[\arXiv{0802.1583} [astro-ph]].

\bibitem{Mandal:2020}
S.~Mandal, D.~Wang and P.~K.~Sahoo,
``Cosmography in $f(Q)$ gravity'',\\
Phys. Rev. D \textbf{102} (2020), 124029
\doi{10.1103/PhysRevD.102.124029}
[\arXiv{2011.00420} [gr-qc]].

\bibitem{Capozziello:2020}
S.~Capozziello, R.~D'Agostino and O.~Luongo,
``High-redshift cosmography: auxiliary variables versus Pad{\'e} polynomials'',
Mon. Not. Roy. Astron. Soc. \textbf{494} (2020) no.2, 2576-2590
\doi{10.1093/mnras/staa871}
[\arXiv{2003.09341} [astro-ph.CO]].

\bibitem{Visser:2004}
M.~Visser,
``Cosmography: Cosmology without the Einstein equations'',\\
Gen. Rel. Grav. \textbf{37} (2005), 1541-1548
\doi{10.1007/s10714-005-0134-8}
[\arXiv{gr-qc/0411131} [gr-qc]].

\bibitem{Shafieloo:2012}
A.~Shafieloo, A.~G.~Kim and E.~V.~Linder,
``Gaussian Process Cosmography'',\\
Phys. Rev. D \textbf{85} (2012), 123530
\doi{10.1103/PhysRevD.85.123530}\\{}
[\arXiv{1204.2272} [astro-ph.CO]].

\bibitem{Sathyaprakash:2009}
B.~S.~Sathyaprakash, B.~F.~Schutz and C.~Van Den Broeck,\\
``Cosmography with the Einstein Telescope'',
Class. Quant. Grav. \textbf{27} (2010), 215006
\doi{10.1088/0264-9381/27/21/215006}
[\arXiv{0906.4151} [astro-ph.CO]].

\bibitem{Aviles:2012a}
A.~Aviles, C.~Gruber, O.~Luongo and H.~Quevedo,
``Cosmography and constraints on the equation of state of the Universe in various parametrizations'',\\
Phys. Rev. D \textbf{86} (2012), 123516
\doi{10.1103/PhysRevD.86.123516}\\{}
[\arXiv{1204.2007} [astro-ph.CO]].

\bibitem{Lavaux:2011}
G.~Lavaux and B.~D.~Wandelt,
``Precision cosmography with stacked voids'',
Astrophys. J. \textbf{754} (2012), 109
\doi{10.1088/0004-637X/754/2/109}\\{}
[\arXiv{1110.0345} [astro-ph.CO]].

\bibitem{Dunsby:2015}
P.~K.~S.~Dunsby and O.~Luongo,
``On the theory and applications of modern cosmography'',
Int. J. Geom. Meth. Mod. Phys. \textbf{13} (2016) no.03, 1630002
\doi{10.1142/S0219887816300026}
[\arXiv{1511.06532} [gr-qc]].

\bibitem{Courtois:2013}
H.~M.~Courtois, D.~Pomarede, R.~B.~Tully and D.~Courtois,
``Cosmography of the Local Universe'',
Astron. J. \textbf{146} (2013), 69
\doi{10.1088/0004-6256/146/3/69}
[\arXiv{1306.0091} [astro-ph.CO]].

\bibitem{Cattoen:2008}
C.~Catt\"oen and M.~Visser,
``Cosmographic Hubble fits to the supernova data'',\\
Phys. Rev. D \textbf{78} (2008), 063501
\doi{10.1103/PhysRevD.78.063501}\\{}
[\arXiv{0809.0537} [gr-qc]].

\bibitem{Capozziello:2008GRB}
S.~Capozziello and L.~Izzo,
``Cosmography by GRBs'',\\
Astron. Astrophys. \textbf{490} (2008), 31
\doi{10.1051/0004-6361:200810337}
[\arXiv{0806.1120} [astro-ph]].


\bibitem{Vitagliano:2009}
V.~Vitagliano, J.~Q.~Xia, S.~Liberati and M.~Viel,
``High-Redshift Cosmography'',
JCAP \textbf{03} (2010), 005
\doi{10.1088/1475-7516/2010/03/005}\\{}
[\arXiv{0911.1249} [astro-ph.CO]].

\bibitem{Cattoen:2007}
C.~Catt\"oen and M.~Visser,\\
``Cosmography: Extracting the Hubble series from the supernova data'',\\{}
[\arXiv{gr-qc/0703122} [gr-qc]].

\bibitem{Xia:2011}
J.~Q.~Xia, V.~Vitagliano, S.~Liberati and M.~Viel,\\
``Cosmography beyond standard candles and rulers'',\\
Phys. Rev. D \textbf{85} (2012), 043520
\doi{10.1103/PhysRevD.85.043520}\\{}
[\arXiv{1103.0378} [astro-ph.CO]].

\bibitem{Aviles:2012b}
A.~Aviles, A.~Bravetti, S.~Capozziello and O.~Luongo,
``Updated constraints on $f(R)$ gravity from cosmography'',
Phys. Rev. D \textbf{87} (2013) no.4, 044012
\doi{10.1103/PhysRevD.87.044012}
[\arXiv{1210.5149} [gr-qc]].

\bibitem{Capozziello:2017ddd}
S.~Capozziello, R.~D'Agostino and O.~Luongo,\\
``Rational approximations of $f(R)$ cosmography through Pad{\'e} polynomials'',\\
JCAP \textbf{05} (2018), 008
\doi{10.1088/1475-7516/2018/05/008}
[\arXiv{1709.08407} [gr-qc]].

\bibitem{Luongo:2011zz}
O.~Luongo,
``Cosmography with the Hubble parameter'',\\
Mod. Phys. Lett. A \textbf{26} (2011), 1459-1466
\doi{10.1142/S0217732311035894}


\bibitem{Lobo:2020}
F.~S.~N.~Lobo, J.~P.~Mimoso and M.~Visser,\\
``Cosmographic analysis of redshift drift'',\\
JCAP \textbf{04} (2020), 043
\doi{10.1088/1475-7516/2020/04/043}
[\arXiv{2001.11964} [gr-qc]].

\bibitem{Visser:2009}
M.~Visser and C.~Catt\"oen,
``Cosmographic analysis of dark energy'',
\doi{10.1142/9789814293792{\_}0022}
[\arXiv{0906.5407} [gr-qc]].

\clearpage
\bibitem{Heinesen:2021}
A.~Heinesen,\\
``Redshift drift cosmography for model-independent cosmological inference'',\\
Phys. Rev. D \textbf{104} (2021) no.12, 123527
\doi{10.1103/PhysRevD.104.123527}
[\arXiv{2107.08674} [astro-ph.CO]].

\bibitem{Apostolopoulos:2016}
Apostolopoulos, P.S.  ``Spatially inhomogeneous and irrotational geometries admitting Intrinsic Conformal Symmetries''.   
\emph{Phys. Rev. D} \textbf{2016}, \emph{94}, 124052. 
\doi{10.1103/PhysRevD.94.124052}. 
[\arXiv{1612.01853} [gr-qc]].

\bibitem{Apostolopoulos:2024}
Apostolopoulos, P.S.; Naidoo, N.  ``Inhomogeneous brane models''. \\ 
\emph{Gen. Rel. Grav.} \textbf{2025}, \emph{57}, 1. 
\do{10.1007/s10714-024-03337-2}. 
[\arXiv{2404.08929} [gr-qc]].

\bibitem{Blandford:2004}
R.~D.~Blandford, M.~A.~Amin, E.~A.~Baltz, K.~Mandel and P.~J.~Marshall,
``Cosmokinetics'',
ASP Conf. Ser. \textbf{339} (2005), 27
[\arXiv{astro-ph/0408279} [astro-ph]].

\bibitem{Nair:2011}
R.~Nair, S.~Jhingan and D.~Jain,
``Cosmokinetics: A joint analysis of Standard Candles, Rulers and Cosmic Clocks'',
JCAP \textbf{01} (2012), 018
\doi{10.1088/1475-7516/2012/01/018}
[\arXiv{1109.4574} [astro-ph.CO]].

\bibitem{Shapiro:2005}
C.~Shapiro and M.~S.~Turner,
``What do we really know about cosmic acceleration?'',
Astrophys. J. \textbf{649} (2006), 563-569
\doi{10.1086/506470}\\{}
[\arXiv{astro-ph/0512586} [astro-ph]].

\bibitem{Linder:2008}
E.~V.~Linder,
``Mapping the Cosmological Expansion'',\\
Rept. Prog. Phys. \textbf{71} (2008), 056901
\doi{10.1088/0034-4885/71/5/056901}
[\arXiv{0801.2968} [astro-ph]].

\bibitem{Cattoen:2007xx}
C.~Catt\"oen and M.~Visser,
``Cosmodynamics: Energy conditions, Hubble bounds, density bounds, time and distance bounds'',\\
Class. Quant. Grav. \textbf{25} (2008), 165013
\doi{10.1088/0264-9381/25/16/165013}
[\arXiv{0712.1619} [gr-qc]].


\bibitem{Visser:2003jerk}
M.~Visser,
``Jerk and the cosmological equation of state'',\\
Class. Quant. Grav. \textbf{21} (2004), 2603-2616
\doi{10.1088/0264-9381/21/11/006}
[\arXiv{gr-qc/0309109} [gr-qc]].

\medskip
\hrule\hrule\hrule



\bibitem{Synge:1961}
 John~Lighton ~Synge, ``Relativity: the General Theory'',\\
  (North-Holland, Amsterdam, 1961).
 
 \enlargethispage{20pt} 
  \bibitem{Ellis:2023}
G.~F.~R.~Ellis and D.~Garfinkle,\\
``The Synge G-Method: cosmology, wormholes, firewalls, geometry'',\\
Class. Quant. Grav. \textbf{41} (2024) no.7, 077002
\doi{10.1088/1361-6382/ad2f14}
[\arXiv{2311.06881} [gr-qc]].



\bibitem{Morris-Thorne}
 M. S. Morris and K. S. Thorne, 
 ``Wormholes in spacetime and their use for interstellar travel: 
 A tool for teaching general relativity'', \\
 American Journal of Physics 56 (1988) 395--412.
\doi{10.1119/1.15620}

 \bibitem{MTY}
M.~S.~Morris, K.~S.~Thorne and U.~Yurtsever,\\
``Wormholes, Time Machines, and the Weak Energy Condition'',\\
Phys. Rev. Lett. \textbf{61} (1988), 1446--1449.
\doi{10.1103/PhysRevLett.61.1446}





\medskip
\hrule\hrule\hrule



\clearpage
\bibitem{Visser:1989a}
M.~Visser,
``Traversable wormholes: Some simple examples'',\\
Phys. Rev. D \textbf{39} (1989), 3182-3184.\\
\doi{10.1103/PhysRevD.39.3182}
[\arXiv{0809.0907} [gr-qc]].

\bibitem{Visser:1989b}
M.~Visser,\\
``Traversable wormholes from surgically modified Schwarzschild space-times'',\\
Nucl. Phys. B \textbf{328} (1989), 203-212.\\
\doi{10.1016/0550-3213(89)90100-4}
[\arXiv{0809.0927} [gr-qc]].

\bibitem{Hochberg:1990}
D.~Hochberg,
``Lorentzian wormholes in higher order gravity theories'',\\
Phys. Lett. B \textbf{251} (1990), 349-354.\\
\doi{10.1016/0370-2693(90)90718-L}

\bibitem{Frolov:1990}
V.~P.~Frolov and I.~D.~Novikov,\\
``Physical Effects in Wormholes and Time Machine'',\\
Phys. Rev. D \textbf{42} (1990), 1057-1065.
\doi{10.1103/PhysRevD.42.1057}

\bibitem{Roman:1992}
T.~A.~Roman,
``Inflating Lorentzian wormholes'',
Phys. Rev. D \textbf{47} (1993), 1370-1379\\{}
\doi{10.1103/PhysRevD.47.1370}
[\arXiv{gr-qc/9211012} [gr-qc]].

\bibitem{Hochberg:1992}
D.~Hochberg and T.~W.~Kephart,
``Wormhole cosmology and the horizon problem'',
Phys. Rev. Lett. \textbf{70} (1993), 2665-2668\\
\doi{10.1103/PhysRevLett.70.2665}
[\arXiv{gr-qc/9211006} [gr-qc]].





\bibitem{Cramer:1994}
J.~G.~Cramer, R.~L.~Forward, M.~S.~Morris, M.~Visser, G.~Benford and G.~A.~Landis,
``Natural wormholes as gravitational lenses'',
Phys. Rev. D \textbf{51} (1995), 3117-3120.
\doi{10.1103/PhysRevD.51.3117}
[\arXiv{astro-ph/9409051} [astro-ph]].

\bibitem{Visser:1995}
M.~Visser,
``Lorentzian wormholes: From Einstein to Hawking'',\\
AIP Press (now Springer--Verlag), 1995. 

\bibitem{Poisson:1995}
E.~Poisson and M.~Visser,
``Thin shell wormholes: Linearization stability'',\\
Phys. Rev. D \textbf{52} (1995), 7318-7321.\\
\doi{10.1103/PhysRevD.52.7318}
[\arXiv{gr-qc/9506083} [gr-qc]].

\bibitem{Kar:1994}
S.~Kar,
``Evolving wormholes and the weak energy condition'',\\
Phys. Rev. D \textbf{49} (1994), 862-865
\doi{10.1103/PhysRevD.49.862}


\bibitem{Kar:1995}
S.~Kar and D.~Sahdev,
``Evolving Lorentzian wormholes'',\\
Phys. Rev. D \textbf{53} (1996), 722-730
\doi{10.1103/PhysRevD.53.722}
[\arXiv{gr-qc/9506094} [gr-qc]].

\bibitem{Hochberg:1995}
D.~Hochberg,
``Quantum mechanical Lorentzian wormholes in cosmological backgrounds'',
Phys. Rev. D \textbf{52} (1995), 6846-6855
\doi{10.1103/PhysRevD.52.6846}
[\arXiv{gr-qc/9507001} [gr-qc]].



\bibitem{Hochberg:1997}
D.~Hochberg and M.~Visser,\\
``Geometric structure of the generic static traversable wormhole throat'',\\
Phys. Rev. D \textbf{56} (1997), 4745-4755.\\
\doi{10.1103/PhysRevD.56.4745}
[\arXiv{gr-qc/9704082} [gr-qc]].

\clearpage
\bibitem{Visser:1997}
M.~Visser and D.~Hochberg,
``Generic wormhole throats'',\\
Annals Israel Phys. Soc. \textbf{13} (1997), 249.\\{}
[\arXiv{gr-qc/9710001} [gr-qc]].

\bibitem{Hochberg:1998}
D.~Hochberg and M.~Visser,\\
``General dynamic wormholes and violation of the null energy condition'',\\{}
[\arXiv{gr-qc/9901020} [gr-qc]].

\bibitem{Krasnikov:1999}
S.~Krasnikov,
``A Traversable wormhole'',
Phys. Rev. D \textbf{62} (2000), 084028
[erratum: Phys. Rev. D \textbf{76} (2007), 109902]
\doi{10.1103/PhysRevD.76.109902}
[\arXiv{gr-qc/9909016} [gr-qc]].


\bibitem{Armendariz-Picon:2002}
C.~Armendariz-Picon,
``On a class of stable, traversable Lorentzian wormholes in classical general relativity'',\\
Phys. Rev. D \textbf{65} (2002), 104010
\doi{10.1103/PhysRevD.65.104010}
[\arXiv{gr-qc/0201027} [gr-qc]].



\bibitem{Lemos:2003}
J.~P.~S.~Lemos, F.~S.~N.~Lobo and S.~Quinet de Oliveira,\\
``Morris--Thorne wormholes with a cosmological constant'',\\
Phys. Rev. D \textbf{68} (2003), 064004.\\
\doi{10.1103/PhysRevD.68.064004}
[\arXiv{gr-qc/0302049} [gr-qc]].

\bibitem{Visser:2003}
M.~Visser, S.~Kar and N.~Dadhich,\\
``Traversable wormholes with arbitrarily small energy condition violations'',\\
Phys. Rev. Lett. \textbf{90} (2003), 201102.\\
\doi{10.1103/PhysRevLett.90.201102}
[\arXiv{gr-qc/0301003} [gr-qc]].

\enlargethispage{20pt}
\bibitem{Kar:2004}
S.~Kar, N.~Dadhich and M.~Visser,\\
``Quantifying energy condition violations in traversable wormholes'',\\
Pramana \textbf{63} (2004), 859-864.\\
\doi{10.1007/BF02705207}
[\arXiv{gr-qc/0405103} [gr-qc]].


\bibitem{Lobo:2005}
F.~S.~N.~Lobo,
``Phantom energy traversable wormholes'',\\
Phys. Rev. D \textbf{71} (2005), 084011.\\
\doi{10.1103/PhysRevD.71.084011}
[\arXiv{gr-qc/0502099} [gr-qc]].

\bibitem{Sushkov:2005}
S.~V.~Sushkov,
``Wormholes supported by a phantom energy'',\\
Phys. Rev. D \textbf{71} (2005), 043520.\\
\doi{10.1103/PhysRevD.71.043520}
[\arXiv{gr-qc/0502084} [gr-qc]].

\bibitem{Harko:2013}
T.~Harko, F.~S.~N.~Lobo, M.~K.~Mak and S.~V.~Sushkov,
``Modified-gravity wormholes without exotic matter'',
Phys. Rev. D \textbf{87} (2013) no.6, 067504
\doi{10.1103/PhysRevD.87.067504}
[\arXiv{1301.6878} [gr-qc]].

\bibitem{Damour:2007}
T.~Damour and S.~N.~Solodukhin,
``Wormholes as black hole foils'',\\
Phys. Rev. D \textbf{76} (2007), 024016.\\
\doi{10.1103/PhysRevD.76.024016}
[\arXiv{0704.2667} [gr-qc]].

\bibitem{Lobo:2007}
F.~S.~N.~Lobo, 
``Exotic solutions in General Relativity: Traversable wormholes and `warp drive' spacetimes'',
[\arXiv{0710.4474} [gr-qc]].

\clearpage
\bibitem{Martin-Moruno:2009}
P.~Mart\'i{}n--Moruno and P.~F.~Gonzalez-Diaz,
``Lorentzian wormholes: Evaporating a time machine!'',
\doi{10.1142/9789814374552{\_}0158}


\bibitem{Konoplya:2010}
R.~A.~Konoplya and A.~Zhidenko,
``Passage of radiation through wormholes of arbitrary shape'',
Phys. Rev. D \textbf{81} (2010), 124036
\doi{10.1103/PhysRevD.81.124036}
[\arXiv{1004.1284} [hep-th]].



\bibitem{Nakajima:2012}
K.~Nakajima and H.~Asada,\\
``Deflection angle of light in an Ellis wormhole geometry'',\\
Phys. Rev. D \textbf{85} (2012), 107501.\\
\doi{10.1103/PhysRevD.85.107501}
[\arXiv{1204.3710} [gr-qc]].


\bibitem{Lobo:2017}
F.~S.~N.~Lobo,
``Wormholes, Warp Drives and Energy Conditions'',\\
Fundam. Theor. Phys. \textbf{189} (2017), pp.-279
Springer, 2017,\\
ISBN 978-3-319-55181-4, 978-3-319-85588-2, 978-3-319-55182-1
\doi{10.1007/978-3-319-55182-1}
[\arXiv{2103.05610} [gr-qc]].

\bibitem{Roman:2004}
T.~A.~Roman,
``Some thoughts on energy conditions and wormholes'',
\doi{10.1142/9789812704030{\_}0236}\\{}
[\arXiv{gr-qc/0409090} [gr-qc]].




\bibitem{Boonserm:2018}
P.~Boonserm, T.~Ngampitipan, A.~Simpson and M.~Visser\\
``Exponential metric represents a traversable wormhole'',\\
Phys. Rev. D \textbf{98} (2018) no.8, 084048.\\
\doi{10.1103/PhysRevD.98.084048}
[\arXiv{1805.03781} [gr-qc]].

\bibitem{DuttaRoy:2019}
P.~Dutta Roy, S.~Aneesh and S.~Kar,
``Revisiting a family of wormholes: geometry, matter, scalar quasinormal modes and echoes'',
Eur. Phys. J. C \textbf{80} (2020) no.9, 850
\doi{10.1140/epjc/s10052-020-8409-5}
[\arXiv{1910.08746} [gr-qc]].

\bibitem{Kar:2022}
S.~Kar, S.~Bose and S.~Aneesh,
``Towards constraining realistic Lorentzian wormholes through observations'',\\
\doi{10.1142/9789811258251{\_}0071}







\bibitem{Alcubierre:1994}
M.~Alcubierre,
``The Warp drive: Hyperfast travel within general relativity'',
Class. Quant. Grav. \textbf{11} (1994), L73-L77
\doi{10.1088/0264-9381/11/5/001}
[\arXiv{gr-qc/0009013} [gr-qc]].

\bibitem{Ford:2000}
L.~H.~Ford and T.~A.~Roman,
``Negative energy, wormholes and warp drive'',
Sci. Am. \textbf{282N1} (2000), 30-37

\bibitem{Lobo:2004-warp}
F.~S.~N.~Lobo and M.~Visser,
``Fundamental limitations on 'warp drive' spacetimes'',
Class. Quant. Grav. \textbf{21} (2004), 5871-5892
\doi{10.1088/0264-9381/21/24/011}
[\arXiv{gr-qc/0406083} [gr-qc]].

\bibitem{Everett:1995}
A.~E.~Everett,
``Warp drive and causality'',
Phys. Rev. D \textbf{53} (1996), 7365-7368
\doi{10.1103/PhysRevD.53.7365}


\bibitem{Everett:1997}
A.~E.~Everett and T.~A.~Roman,
``A Superluminal subway: The Krasnikov tube'',
Phys. Rev. D \textbf{56} (1997), 2100-2108
\doi{10.1103/PhysRevD.56.2100}
[\arXiv{gr-qc/9702049} [gr-qc]].

\bibitem{Pfenning:1997}
M.~J.~Pfenning and L.~H.~Ford,
``The Unphysical nature of 'warp drive''',
Class. Quant. Grav. \textbf{14} (1997), 1743-1751\\
\doi{10.1088/0264-9381/14/7/011}
[\arXiv{gr-qc/9702026} [gr-qc]].


\bibitem{Clark:1999}
C.~Clark, W.~A.~Hiscock and S.~L.~Larson,
``Null geodesics in the Alcubierre warp drive space-time: The View from the bridge'',
Class. Quant. Grav. \textbf{16} (1999), 3965-3972
\doi{10.1088/0264-9381/16/12/313}
[\arXiv{gr-qc/9907019} [gr-qc]].





\bibitem{Natario:2001}
J.~Nat{\'a}rio,
``Warp drive with zero expansion'',\\
Class. Quant. Grav. \textbf{19} (2002), 1157-1166
\doi{10.1088/0264-9381/19/6/308}
[\arXiv{gr-qc/0110086} [gr-qc]].

\bibitem{Lobo:2002}
F.~Lobo and P.~Crawford,
``Weak energy condition violation and superluminal travel'',
Lect. Notes Phys. \textbf{617} (2003), 277-291
\doi{10.1007/3-540-36973-2{\_}15}
[\arXiv{gr-qc/0204038} [gr-qc]].



\bibitem{Hiscock:1997}
W.~A.~Hiscock,
``Quantum effects in the Alcubierre warp drive space-time'',\\
Class. Quant. Grav. \textbf{14} (1997), L183-L188
\doi{10.1088/0264-9381/14/11/002}
[\arXiv{gr-qc/9707024} [gr-qc]].

\bibitem{Finazzi:2009}
S.~Finazzi, S.~Liberati and C.~Barcel\'o,\\
``Semiclassical instability of dynamical warp drives'',\\
Phys. Rev. D \textbf{79} (2009), 124017
\doi{10.1103/PhysRevD.79.124017}\\{}
[\arXiv{0904.0141} [gr-qc]].

\bibitem{Barcelo:2010-F}
C.~Barcel{\'o}, S.~Finazzi and S.~Liberati,
``Semiclassical instability of warp drives'',
J. Phys. Conf. Ser. \textbf{229} (2010), 012018
\doi{10.1088/1742-6596/229/1/012018}




\bibitem{Santiago:2021-warp}
J.~Santiago, S.~Schuster and M.~Visser,\\
``Generic warp drives violate the null energy condition'',\\
Phys. Rev. D \textbf{105} (2022) no.6, 064038
\doi{10.1103/PhysRevD.105.064038}
[\arXiv{2105.03079} [gr-qc]].

\bibitem{Lobo:2004-linearized}
F.~S.~N.~Lobo and M.~Visser,
``Linearized warp drive and the energy conditions'',
[\arXiv{gr-qc/0412065} [gr-qc]].

\bibitem{Coutant:2011}
A.~Coutant, S.~Finazzi, S.~Liberati and R.~Parentani,\\
``Impossibility of superluminal travel in Lorentz violating theories'',\\
Phys. Rev. D \textbf{85} (2012), 064020
\doi{10.1103/PhysRevD.85.064020}\\{}
[\arXiv{1111.4356} [gr-qc]].

\bibitem{Shoshany:2019}
B.~Shoshany,
``Lectures on Faster-than-Light Travel and Time Travel'',\\
SciPost Phys. Lect. Notes \textbf{10} (2019), 1\\
\doi{10.21468/SciPostPhysLectNotes.10}
[\arXiv{1907.04178} [gr-qc]].

\bibitem{Shoshany:2023}
B.~Shoshany and B.~Snodgrass,
``Warp drives and closed timelike curves'',\\
Class. Quant. Grav. \textbf{41} (2024) no.20, 205005\\
\doi{10.1088/1361-6382/ad74d1}
[\arXiv{2309.10072} [gr-qc]].


\clearpage


\bibitem{Alcubierre:2017-basics}
M.~Alcubierre and F.~S.~N.~Lobo,
``Warp Drive Basics'',\\
Fundam. Theor. Phys. \textbf{189} (2017), 257-279
\doi{10.1007/978-3-319-55182-1{\_}11}

\bibitem{Barcelo:2022-cpc}
C.~Barcel{\'o}, J.~E.~S{\'a}nchez, G.~Garc{\'\i}a-Moreno and G.~Jannes,\\
``Chronology protection implementation in analogue gravity'',\\
Eur. Phys. J. C \textbf{82} (2022) no.4, 299
\doi{10.1140/epjc/s10052-022-10275-3}
[\arXiv{2201.11072} [gr-qc]].

\bibitem{Schuster:2022-adm}
S.~Schuster, J.~Santiago and M.~Visser,
``ADM mass in warp drive spacetimes'',\\
Gen. Rel. Grav. \textbf{55} (2023) no.1, 14
\doi{10.1007/s10714-022-03061-9}
[\arXiv{2205.15950} [gr-qc]].

\bibitem{Liberati:2016-mess}
S.~Liberati,
``Do not mess with time: Probing faster than light travel and chronology protection with superluminal warp drives'',
\doi{10.1142/9789813226609{\_}0120}
[\arXiv{1601.00785} [gr-qc]].

\bibitem{Barcelo:2022-aero}
C.~Barcel{\'o}, V.~Boyanov, L.~J.~Garay, E.~Mart{\'\i}n-Mart{\'\i}nez and J.~M.~S.~Vel{\'a}zquez,
``Warp drive aerodynamics'',
JHEP \textbf{08} (2022), 288
\doi{10.1007/JHEP08(2022)288}
[\arXiv{2207.06458} [gr-qc]].

\bibitem{Barcelo:2010-impossible}
C.~Barcel\'o, S.~Finazzi and S.~Liberati,\\
``On the Impossibility of Superluminal Travel: The Warp Drive Lesson'',\\{}
[\arXiv{1001.4960} [gr-qc]].

\bibitem{Finazzi:2010}
S.~Finazzi, S.~Liberati and C.~Barcel{\'o},\\
``Superluminal warp drives are semiclassically unstable'',\\
J. Phys. Conf. Ser. \textbf{222} (2010), 012046
\doi{10.1088/1742-6596/222/1/012046}

\bibitem{Schuster:2023-frenemies}
S.~Schuster,
``Frenemies with Physicality: Manufacturing Manifold Metrics'',
[\arXiv{2305.08725} [gr-qc]].

\bibitem{Clough:2024}
K.~Clough, T.~Dietrich and S.~Khan,\\
``What no one has seen before: gravitational waveforms from 
warp drive collapse'',\\
The Open Journal of Astrophysics {\bf7} (2014, July).
\doi{10.33232/001c.121868}
[\arXiv{2406.02466} [gr-qc]].


\bibitem{Santiago:2021-tractor}
J.~Santiago, S.~Schuster and M.~Visser,\\
``Tractor Beams, Pressor Beams and Stressor Beams in General Relativity'',\\
Universe \textbf{7} (2021) no.8, 271
\doi{10.3390/universe7080271}
[\arXiv{2106.05002} [gr-qc]].

\bibitem{Visser:2021-tractor}
M.~Visser, J.~Santiago and S.~Schuster,
``Tractor beams, pressor beams, and stressor beams within the context of general relativity'',\\
\doi{10.1142/9789811269776{\_}0063}
[\arXiv{2110.14926} [gr-qc]].

\bibitem{Tippett:2013}
B.~K.~Tippett and D.~Tsang,
``The Blue Box White Paper'',\\{}
[\arXiv{1310.7983} [physics.pop-ph]].


\bibitem{Hiscock:2002}
W.~A.~Hiscock,
``From wormholes to the warp drive: Using theoretical physics to place ultimate bounds on technology'',
[\arXiv{physics/0211114} [physics]].

\bibitem{Obousy:2008}
R.~K.~Obousy and G.~Cleaver,
``Putting the 'Warp' into Warp Drive'',
[\arXiv{0807.1957} [physics.pop-ph]].


\clearpage
\bibitem{Schwarzschild-interior}
Karl Schwarzschild,\\
``Über  das Gravitationsfeld einer Kugel aus inkompressibler Fl\"ussigkeit nach der Einsteinschen Theorie" \\
{}\leftline{\qquad [On the gravitational field of a ball of incompressible fluid following Einstein's theory].}
\leftline{\qquad Sitzungsberichte der Königlich-Preussischen Akademie der Wissenschaften {\bf 7} (2016) 424--434.}


\bibitem{Schwarzschild-exterior}
Karl Schwarzschild,  \\
``Über das Gravitationsfeld eines Massenpunktes nach der Einsteinschen Theorie". \\
{}[On the gravitational field of a point mass following Einstein's theory].
\\ 
\leftline{\qquad Sitzungsberichte der Königlich Preussischen Akademie der Wissenschaften {\bf 7} (1916) 189–-196.}



\bibitem{Droste:1916}
 Johannes Droste,
 ``The field of a single centre in Einstein's theory of gravitation, and the motion of a particle in that field'',  \\
 Proceedings of the Royal Netherlands Academy of Arts and Science. {\bf 19 \# 1} (1917) 197–-215.
(Submitted 27 May 1916.)

\bibitem{Droste:2002a}
Johannes Droste, ``Golden Oldie: 
The field of a single centre in Einstein's theory of gravitation, and the motion of a particle in that field'', \\
General Relativity and Gravitation {\bf 34} (2002) 1545–1563. 
\doi{10.1023/A:1020747322668}


\enlargethispage{30pt}
\bibitem{Droste:2002b}
Johannes Droste, 
``Editor's note: The field of a single centre in Einstein's theory of gravitation, and the motion of a particle in that field'', \\
General Relativity and Gravitation {\bf 34} (2002) 1541–1543. 
\doi{10.1023/A:1020795205829}

\bibitem{Hilbert:1916}
David Hilbert, ``Die Grundlagen der Physik'', [The foundations of Physics],\\
Nachr. Ges. Wiss. G\"ottingen, Math. Phys. Kl. (1917) 53.\\ 
(Submitted 23 December 1916.)

\bibitem{TOV}
 J.~R.~Oppenheimer and  G.~B.~Volkov, (1939). 
 ``On massive neutron cores". \\
 Phys. Rev. {\bf 55 \#4}  (1939)  374--381.  \doi{10.1103/PhysRev.55.374}.
 
 \bigskip
 \hrule\hrule\hrule
 \bigskip

\bibitem{Delgaty:1998}
M.~S.~R.~Delgaty and K.~Lake,
``Physical acceptability of isolated, static, spherically symmetric, perfect fluid solutions of Einstein's equations'',\\
Comput. Phys. Commun. \textbf{115} (1998), 395-415
\doi{10.1016/S0010-4655(98)00130-1}
[\arXiv{gr-qc/9809013} [gr-qc]].

\clearpage
\bibitem{Stephani:2003}
H.~Stephani, D.~Kramer, M.~A.~H.~MacCallum, C.~Hoenselaers 
and E.~Herlt,\\
``Exact solutions of Einstein's field equations'',
Cambridge Univ. Press, 2003,\\
ISBN 978-0-521-46702-5, 978-0-511-05917-9
\doi{10.1017/CBO9780511535185}

 \bibitem{MacCallum:2006}
M.~A.~H.~MacCallum,
``Finding and using exact solutions of the Einstein equations'',
AIP Conf. Proc. \textbf{841} (2006) no.1, 129-143
\doi{10.1063/1.2218172}
[\arXiv{gr-qc/0601102} [gr-qc]].

\bibitem{Griffiths:2009}
J.~B.~Griffiths and J.~Podolsky,\\
``Exact Space-Times in Einstein's General Relativity'',\\
Cambridge University Press, 2009,
ISBN 978-1-139-48116-8
\doi{10.1017/CBO9780511635397}

\bibitem{Hawking-Ellis}
S.~W.~Hawking and G.~F.~R.~Ellis,
``The Large Scale Structure of Space-Time'',
Cambridge University Press, 2023,\\
ISBN 978-1-009-25316-1, 978-1-009-25315-4, 978-0-521-20016-5, 978-0-521-09906-6, 978-0-511-82630-6, 978-0-521-09906-6\\
\doi{10.1017/9781009253161}

\bibitem{Finch-Skea}
 M. R. Finch and J. E. F. Skea, 
 ``A review of the relativistic static fluid sphere'', 
 1998, unpublished.
 
 \bibitem{Tolman:1939}
R.~C.~Tolman,
``Static solutions of Einstein's field equations for spheres of fluid'',
Phys. Rev. \textbf{55} (1939), 364-373\\
\doi{10.1103/PhysRev.55.364}

 \bigskip
 \hrule\hrule\hrule
 \bigskip

\bibitem{Barraco:2002}
D.~Barraco and V.~H.~Hamity,
``Maximum mass of a spherically symmetric isotropic star'',
Phys. Rev. D \textbf{65} (2002), 124028\\
\doi{10.1103/PhysRevD.65.124028}

\bibitem{Mak:2013}
M.~K.~Mak and T.~Harko,
``Isotropic stars in general relativity'',
Eur. Phys. J. C \textbf{73} (2013), 2585
\doi{10.1140/epjc/s10052-013-2585-5}\\{}
[\arXiv{1309.5123} [gr-qc]].

\bibitem{Lake:2008}
K.~Lake,
``Transforming the Einstein static Universe into physically acceptable static fluid spheres'',
Phys. Rev. D \textbf{77} (2008), 127502
\doi{10.1103/PhysRevD.77.127502}
[\arXiv{0804.3092} [gr-qc]].


\bibitem{Rahman:2001}
S.~Rahman and M.~Visser,
``Space-time geometry of static fluid spheres'',\\
Class. Quant. Grav. \textbf{19} (2002), 935-952
\doi{10.1088/0264-9381/19/5/307}
[\arXiv{gr-qc/0103065} [gr-qc]].

\bibitem{Martin:2003}
D.~Martin and M.~Visser,\\
``Bounds on the interior geometry and pressure profile of 
static fluid spheres'',\\
Class. Quant. Grav. \textbf{20} (2003), 3699-3716
\doi{10.1088/0264-9381/20/16/311}
[\arXiv{gr-qc/0306038} [gr-qc]].


\clearpage
\bibitem{Boonserm:2005}
P.~Boonserm, M.~Visser and S.~Weinfurtner,\\
``Generating perfect fluid spheres in general relativity'',\\
Phys. Rev. D \textbf{71} (2005), 124037
\doi{10.1103/PhysRevD.71.124037}
[\arXiv{gr-qc/0503007} [gr-qc]].




\bibitem{Boonserm:2006}
P.~Boonserm, M.~Visser and S.~Weinfurtner,\\
``Solution generating theorems for the TOV equation'',\\
Phys. Rev. D \textbf{76} (2007), 044024
\doi{10.1103/PhysRevD.76.044024}
[\arXiv{gr-qc/0607001} [gr-qc]].

\bibitem{Boonserm:2005-msc}
P.~Boonserm,
``Some exact solutions in general relativity'', \\
(MSc thesis, Victoria University of Wellington)
[\arXiv{gr-qc/0610149} [gr-qc]].

\bibitem{Boonserm:2007}
P.~Boonserm and M.~Visser,\\
``Buchdahl-like transformations for perfect fluid spheres'',\\
Int. J. Mod. Phys. D \textbf{17} (2008), 135-163
\doi{10.1142/S0218271808011912}
[\arXiv{0707.0146} [gr-qc]].



\bibitem{Boonserm:2007b}
P.~Boonserm and M.~Visser,
``Buchdahl-Like Transformations in General Relativity'',
Thai Journal of Mathematics \textbf{5} (2007) no.Number 2, 209-223

\bibitem{Kinreewong:2016}
A.~Kinreewong, P.~Boonserm and T.~Ngampitipan,\\
``Solution Generating Theorems and Tolman-Oppenheimer-Volkov Equation for Perfect Fluid Spheres in Isotropic Coordinates'',
\doi{10.2991/amsm-16.2016.61}

\bibitem{Boonserm:2019-siri}
P.~Boonserm, T.~Ngampitipan and S.~Boonsiri,
``Solving for Schwarzschild solution using variation of parameters and Frobenius method'',
AIP Conf. Proc. \textbf{2184} (2019), 060019
\doi{10.1063/1.5136451}


\bibitem{Boonserm:2021}
P.~Boonserm, K.~Sansook and T.~Ngampitipan,\\
``Quasinormal modes of perfect fluid spheres'',\\
AIP Conf. Proc. \textbf{2423} (2021), 020007
\doi{10.1063/5.0075377}



\bibitem{Mantica:2024}
C.~A.~Mantica and L.~G.~Molinari,
``Tolman-Oppenheimer-Volkoff equation and static spheres in conformal Killing gravity'',
Phys. Rev. D \textbf{111} (2025) no.6, 064085
\doi{10.1103/PhysRevD.111.064085}
[\arXiv{2409.18663} [gr-qc]].
\bigskip
\hrule\hrule\hrule

\bibitem{Martin-Moruno:2017-Rainich}
P.~Mart{\'\i}n-Moruno and M.~Visser,
``Generalized Rainich conditions, generalized stress-energy conditions, and the Hawking-Ellis classification'',\\
Class. Quant. Grav. \textbf{34} (2017) no.22, 225014
\doi{10.1088/1361-6382/aa9039}
[\arXiv{1707.04172} [gr-qc]].

\bibitem{Martin-Moruno:2018-core}
P.~Mart{\'i}n-Moruno and M.~Visser,
``Essential core of the Hawking{\textendash}Ellis types'',
Class. Quant. Grav. \textbf{35} (2018) no.12, 125003
\doi{10.1088/1361-6382/aac147}
[arXiv:1802.00865 [gr-qc]].

\bigskip
\hrule\hrule\hrule


\bibitem{Curiel:2014}
E.~Curiel,
``A Primer on Energy Conditions'',\\
Einstein Stud. \textbf{13} (2017), 43-104.\\
\doi{10.1007/978-1-4939-3210-8{\_}3}
[\arXiv{1405.0403} [physics.hist-ph]].


\bibitem{Martin-Moruno:2017}
P.~Mart\'in-Moruno and M.~Visser,
``Classical and semi-classical energy conditions'',\\
Fundam. Theor. Phys. \textbf{189} (2017), 193-213.\\
\doi{10.1007/978-3-319-55182-1{\_}9}
[\arXiv{1702.05915} [gr-qc]].



\bibitem{Barcelo:2002}
C.~Barcel\'o and M.~Visser,
``Twilight for the energy conditions?'',\\
Int. J. Mod. Phys. D \textbf{11} (2002), 1553-1560.\\
\doi{10.1142/S0218271802002888}
[\arXiv{gr-qc/0205066} [gr-qc]].


\bibitem{Borissova:2025a}
J.~Borissova, S.~Liberati and M.~Visser,\\
``Violations of the null convergence condition in kinematical transitions between singular and regular black holes, horizonless compact objects, and bounces'',\\
Phys. Rev. D \textbf{111} (2025) no.10, 104054.
\doi{10.1103/PhysRevD.111.104054}
[\arXiv{2502.00548} [gr-qc]].

\bibitem{Borissova:2025b}
J.~Borissova, S.~Liberati and M.~Visser, \\
``Timelike convergence condition in regular black-hole spacetimes 
with (anti-)de~Sitter core'', Physical Review D {\bf 112} (2025) 104072
\doi{10.1103/rrc9-g1sv}
[\arXiv{2509.08590} [gr-qc]].



\bibitem{defective}
J.~Baines, R.~Gaur and M.~Visser,
``Defect Wormholes Are Defective'',\\
Universe \textbf{9 \# 10} (2023) 452
\doi{10.3390/universe9100452}
[\arXiv{2308.16624} [gr-qc]].

\bibitem{Feng:2023}
J.~C.~Feng,
``Smooth metrics can hide thin shells'',\\
Class. Quant. Grav. \textbf{40 \# 19} (2023) 197002.
\doi{10.1088/1361-6382/acf2de}
[\arXiv{2308.11885} [gr-qc]].

\bibitem{Simmonds:2025}
C.~Simmonds and M.~Visser,
``Traversable Kaluza-Klein wormholes?'',\\
Universe {\bf 11} (2025) 347
\doi{10.3390/universe11100347}
[\arXiv{2508.02824} [gr-qc]].

\bigskip
\hrule\hrule\hrule

%
%

%




%
%

%
%


\bibitem{Kiselev1}
M.~Visser,
``The Kiselev black hole is neither perfect fluid, nor is it quintessence'',
Class. Quant. Grav. \textbf{37} (2020) no.4, 045001.
\doi{10.1088/1361-6382/ab60b8}
[\arXiv{1908.11058} [gr-qc]].

\bibitem{Kiselev2}
P.~Boonserm, T.~Ngampitipan, A.~Simpson and M.~Visser,\\
``Decomposition of the total stress energy for the generalized Kiselev black hole'',
Phys. Rev. D \textbf{101} (2020) no.2, 024022
\doi{10.1103/PhysRevD.101.024022}
[\arXiv{1910.08008} [gr-qc]].

\bibitem{Kiselev0}
V.~V.~Kiselev,
``Quintessence and black holes'',\\
Class. Quant. Grav. \textbf{20} (2003), 1187-1198.
\doi{10.1088/0264-9381/20/6/310}
[\arXiv{gr-qc/0210040} [gr-qc]].

\medskip
\hrule\hrule\hrule


\bibitem{Herrera:2018a}
L.~Herrera,
``New definition of complexity for self-gravitating fluid distributions: The spherically symmetric, static case'',
Phys. Rev. D \textbf{97} (2018) no.4, 044010
\doi{10.1103/PhysRevD.97.044010}
[\arXiv{1801.08358} [gr-qc]].

\bibitem{Herrera:2018b}
L.~Herrera, A.~Di Prisco and J.~Ospino,
``Definition of complexity for dynamical spherically symmetric dissipative self-gravitating fluid distributions'',\\
Phys. Rev. D \textbf{98} (2018) no.10, 104059
\doi{10.1103/PhysRevD.98.104059}
[\arXiv{1811.08364} [gr-qc]].

\bibitem{Herrera:2019a}
L.~Herrera, A.~Di Prisco and J.~Ospino,\\
``Complexity factors for axially symmetric static sources'',\\
Phys. Rev. D \textbf{99} (2019) no.4, 044049
\doi{10.1103/PhysRevD.99.044049}
[\arXiv{1902.10133} [gr-qc]].

\bibitem{Herrera:2019b}
L.~Herrera, A.~Di Prisco and J.~Carot,
``Complexity of the Bondi metric'',\\
Phys. Rev. D \textbf{99} (2019) no.12, 124028
\doi{10.1103/PhysRevD.99.124028}
[\arXiv{1906.08640} [gr-qc]].

\bibitem{Herrera:2020}
L.~Herrera, A.~Di Prisco and J.~Ospino,
``Quasi-homologous evolution of self-gravitating systems with vanishing complexity factor'',\\
Eur. Phys. J. C \textbf{80} (2020) no.7, 631
\doi{10.1140/epjc/s10052-020-8202-5}
[\arXiv{2007.12029} [gr-qc]].

\bibitem{Herrera:2023}
L.~Herrera,
``Complexity and Simplicity of Self{\textendash}Gravitating Fluids'',\\
\doi{10.1007/978-981-97-1172-7{\_}8}
[\arXiv{2304.05870} [gr-qc]].

\bibitem{Herrera:2024}
L.~Herrera and A.~Di Prisco,
``Cracking and complexity of self-gravitating dissipative compact objects'',
Phys. Rev. D \textbf{109} (2024) no.6, 064071
\doi{10.1103/PhysRevD.109.064071}
[\arXiv{2404.04901} [gr-qc]].

\bibitem{Herrera:2025}
L.~Herrera, A.~Di Prisco and J.~Ospino,\\
``Complexity hierarchies in Euclidean stars'',\\
Symmetry \textbf{17} (2025), 1517
[\arXiv{2509.26384} [gr-qc]].

\bibitem{Herrera:2022}
L.~Herrera, A.~Di Prisco and J.~Ospino,
``Non-Static Fluid Spheres Admitting a Conformal Killing Vector: Exact Solutions'',
Universe \textbf{8} (2022) no.6, 296
\doi{10.3390/universe8060296}
[\arXiv{2206.02143} [gr-qc]].

\bibitem{Herrera:2024b}
L.~Herrera, A.~Di Prisco and J.~Ospino,
``The Post-Quasi-Static Approximation: An Analytical Approach to Gravitational Collapse'',
Symmetry \textbf{16} (2024) no.3, 341
\doi{10.3390/sym16030341}
[\arXiv{2403.07550} [gr-qc]].

\bibitem{Herrera:2024c}
L.~Herrera, A.~Di Prisco and J.~Ospino,
``Evolution of Self-Gravitating Fluid Spheres Involving Ghost Stars'', \\
Symmetry \textbf{16} (2024) no.11, 1422
\doi{10.3390/sym16111422}
[\arXiv{2411.04544} [gr-qc]].

\bibitem{Herrera:2025b}
L.~Herrera, A.~Di Prisco and J.~Ospino,
``The Birth of a Ghost Star'',
Entropy \textbf{27} (2025) no.4, 412
\doi{10.3390/e27040412}
[\arXiv{2505.02871} [gr-qc]].





\medskip
\hrule\hrule\hrule

\bibitem{Casadio:2019}
R.~Casadio, E.~Contreras, J.~Ovalle, A.~Sotomayor and Z.~Stuchlick,
``Isotropization and change of complexity by gravitational decoupling'',
Eur. Phys. J. C \textbf{79} (2019) no.10, 826
\doi{10.1140/epjc/s10052-019-7358-3}
[\arXiv{1909.01902} [gr-qc]].

\bibitem{Sharif:2018}
M.~Sharif and I.~I.~Butt,
``Complexity Factor for Charged Spherical System'',
Eur. Phys. J. C \textbf{78} (2018) no.8, 688
\doi{10.1140/epjc/s10052-018-6121-5}
[\arXiv{1808.00903} [gr-qc]].

\clearpage
\bibitem{Yousaf:2020}
Z.~Yousaf, M.~Y.~Khlopov, M.~Z.~Bhatti and T.~Naseer,
``Influence of Modification of Gravity on the Complexity Factor of Static Spherical Structures'',
Mon. Not. Roy. Astron. Soc. \textbf{495} (2020) no.4, 4334-4346
\doi{10.1093/mnras/staa1470}\\{}
[\arXiv{2005.10697} [gr-qc]].

\bibitem{Yousaf:2020b}
Z.~Yousaf,
``Definition of complexity factor for self-gravitating systems in Palatini $f(R)$ gravity'',\\
Phys. Scripta \textbf{95} (2020) no.7, 075307
\doi{10.1088/1402-4896/ab9479}
[\arXiv{2006.01642} [gr-qc]].

\bibitem{Yousaf:2020c}
Z.~Yousaf, M.~Z.~Bhatti and T.~Naseer,
``Measure of complexity for dynamical self-gravitating structures'',\\
Int. J. Mod. Phys. D \textbf{29} (2020) no.09, 2050061
\doi{10.1142/S0218271820500613}

\bibitem{Contreras:2021}
E.~Contreras and E.~Fuenmayor,
``Gravitational cracking and complexity in the framework of gravitational decoupling'',\\
Phys. Rev. D \textbf{103} (2021) no.12, 124065
\doi{10.1103/PhysRevD.103.124065}
[\arXiv{2107.01140} [gr-qc]].

\bibitem{Carrasco-Hidalgo:2021}
M.~Carrasco-Hidalgo and E.~Contreras,
``Ultracompact stars with polynomial complexity by gravitational decoupling'',\\
Eur. Phys. J. C \textbf{81} (2021) no.8, 757
\doi{10.1140/epjc/s10052-021-09557-z}
[\arXiv{2108.10311} [gr-qc]].

\bibitem{Maurya:2022}
S.~K.~Maurya, A.~Errehymy, R.~Nag and M.~Daoud,
``Role of Complexity on Self-gravitating Compact Star by Gravitational Decoupling'',
Fortsch. Phys. \textbf{70} (2022) no.5, 2200041
\doi{10.1002/prop.202200041}

\bibitem{Maurya:2022b}
S.~K.~Maurya, M.~Govender, S.~Kaur and R.~Nag,
``Isotropization of embedding Class I spacetime and anisotropic system generated by complexity factor in the framework of gravitational decoupling'',
Eur. Phys. J. C \textbf{82} (2022) no.2, 100\\
\doi{10.1140/epjc/s10052-022-10030-8}


\bibitem{Arias:2021}
C.~Arias, E.~Contreras, E.~Fuenmayor and A.~Ramos,\\
``Anisotropic star models in the context of vanishing complexity'',\\
Annals Phys. \textbf{436} (2022), 168671
\doi{10.1016/j.aop.2021.16867}\\{}
[\arXiv{2208.10594} [gr-qc]].

\bibitem{Naseer:2024a}
T.~Naseer and M.~Sharif,\\
``Charged anisotropic Starobinsky models admitting vanishing complexity'',\\
Phys. Dark Univ. \textbf{46} (2024), 101595
\doi{10.1016/j.dark.2024.101595}

\bibitem{Sharif:2023}
M.~Sharif and T.~Naseer,\\
``Charged anisotropic models with complexity-free condition'',\\
Annals Phys. \textbf{453} (2023), 169311
\doi{10.1016/j.aop.2023.169311}\\{}
[\arXiv{2306.00328} [gr-qc]].

\bibitem{Nazar:2021}
H.~Nazar, A.~H.~Alkhaldi, G.~Abbas and M.~R.~Shahzad,\\
``Complexity factor for anisotropic self-gravitating sphere in Rastall gravity'',\\
Int. J. Mod. Phys. A \textbf{36} (2021) no.31n32, 2150233
\doi{10.1142/S0217751X2150233X}

\bibitem{Kaur:2023}
S.~Kaur, S.~K.~Maurya and S.~Shukla,\\
``Anisotropic fluid solution in $f(Q)$ gravity satisfying vanishing complexity factor'',\\
Phys. Scripta \textbf{98} (2023) no.10, 105304
\doi{10.1088/1402-4896/acf348}

\bibitem{Naseer:2024b}
T.~Naseer and M.~Sharif,\\
``Role of decoupling and Rastall parameters on Krori{\textendash}Barua and Tolman IV models generated by isotropization and complexity factor'',\\
Class. Quant. Grav. \textbf{41} (2024) no.24, 245006
\doi{10.1088/1361-6382/ad8d9d}

\bibitem{Naseer:2024c}
T.~Naseer and M.~Sharif,
``Extending Finch-Skea isotropic model to anisotropic domain in modified f(R, T) gravity'',\\
Phys. Scripta \textbf{99} (2024) no.7, 075012
\doi{10.1088/1402-4896/ad504c}
[\arXiv{2412.03291} [gr-qc]].

\bibitem{Khan:2024}
S.~Khan and Z.~Yousaf,
``Complexity-free charged anisotropic Finch-Skea model satisfying Karmarkar condition'',\\
Phys. Scripta \textbf{99} (2024) no.5, 055303
\doi{10.1088/1402-4896/ad38e2}




\bibitem{Naseer:2025a}
T.~Naseer, M.~Sharif, M.~Faiza, K.~S.~Nisar and M.~Mahmoud,\\
``Insights of traversable wormhole geometries under complexity factor and different equations of state in modified gravity'',\\
Annals Phys. \textbf{479} (2025), 170076
\doi{10.1016/j.aop.2025.170076}

\bibitem{Naseer:2025b}
T.~Naseer, M.~Sharif, M.~Faiza, F.~Afandi, M.~R.~Eid and A.~H.~Abdel-Aty,\\
``Exploring traversable wormholes in modified theory: Complexity and VIQ perspective under solitonic quantum wave dark matter halo'',\\
Phys. Dark Univ. \textbf{50} (2025), 102112
\doi{10.1016/j.dark.2025.102112}

\bibitem{Harko:2002}
T.~Harko and M.~K.~Mak,
``Anisotropic relativistic stellar models'',
Annalen Phys. \textbf{11} (2002), 3-13
\doi{10.1002/andp.20025140103}
[\arXiv{gr-qc/0302104} [gr-qc]].

\bibitem{Tello-Ortiz:2020}
F.~Tello-Ortiz, M.~Malaver, {\'A}.~Rinc{\'o}n and Y.~Gomez-Leyton,
``Relativistic anisotropic fluid spheres satisfying a non-linear equation of state'',
Eur. Phys. J. C \textbf{80} (2020) no.5, 371
\doi{10.1140/epjc/s10052-020-7956-0}
[\arXiv{2005.11038} [gr-qc]].

\bibitem{Singh:2016}
K.~N.~Singh, P.~Bhar and N.~Pant,
``A new solution of embedding class I representing anisotropic fluid sphere in general relativity'',
Int. J. Mod. Phys. D \textbf{25} (2016) no.14, 1650099
\doi{10.1142/S0218271816500991}
[\arXiv{1604.01013} [gr-qc]].

\medskip
\hrule\hrule\hrule

\bibitem{Boonserm:mimic}
P.~Boonserm, T.~Ngampitipan and M.~Visser,\\
``Mimicking static anisotropic fluid spheres in general relativity'',\\
Int. J. Mod. Phys. D \textbf{25} (2015) no.02, 1650019
\doi{10.1142/S021827181650019X}
[\arXiv{1501.07044} [gr-qc]].

\bigskip
\hrule\hrule\hrule

\end{thebibliography}
\end{document}